%% file: main.tex
\documentclass[runningheads]{comsis2}
\def\journalissue{} 
\def\paperidnum{} 
\usepackage[pagewise]{lineno}

\usepackage{graphicx} 
\usepackage[dvipsnames]{xcolor}
\usepackage{multirow}
\usepackage{hyperref}
\usepackage{pdflscape}
\usepackage{placeins} 

\usepackage[algo2e,linesnumbered,commentsnumbered, lined, boxed,ruled,vlined]{algorithm2e}
\def\HiLi{\leavevmode\rlap{\hbox to \hsize{\color{green!10}\leaders\hrule height .8\baselineskip depth .5ex\hfill}}}
\def\HiLiScalar{\leavevmode\rlap{\hbox to \hsize{\color{blue!10}\leaders\hrule height .8\baselineskip depth .5ex\hfill}}}
\def\HiLiAvx{\leavevmode\rlap{\hbox to \hsize{\color{red!10}\leaders\hrule height .8\baselineskip depth .5ex\hfill}}}

\SetAlFnt{\small}
\SetAlCapFnt{\small}
\SetAlCapNameFnt{\small}
\SetAlCapHSkip{0pt}
\SetKwProg{Fn}{function}{}{}
\SetKwBlock{Begin}{ }{ }

\usepackage{pgfplots}
\pgfplotsset{compat=newest}
\usetikzlibrary{patterns}
\usepackage{subcaption}
\definecolor{color_0}{rgb}{0.3, 0.3, 0.3}
\definecolor{color_x}{rgb}{0.7, 0.75, 0.71}
\definecolor{color_mkl}{rgb}{0.7, 0.75, 0.71}

\definecolor{color_1}{rgb}{0.0, 0.3, 0.8}
\definecolor{color_2}{rgb}{0.8, 0.0, 0.0}
\definecolor{color_3}{rgb}{0.75, 0.0, 0.2}
\definecolor{color_4}{rgb}{0.5, 0.0, 0.1}
\definecolor{color_5}{rgb}{0.0, 0.34, 0.25}
\definecolor{color_6}{rgb}{0.4, 0.7, 0.0}
\definecolor{color_7}{rgb}{0.53, 0.47, 0.76}

\usepackage{textcomp}

\title{SPC5: an efficient SpMV framework vectorized using ARM SVE and x86 AVX-512}
\author{Evann Regnault\inst{1} \and B\'erenger Bramas\inst{2}}

\institute{Strasbourg University\\
  UFR de Mathématique et d’Informatique\\
  7, rue René Descartes\\
  67084 Strasbourg\\
  France\\
  \email{Evann.Regnault@etu.unistra.fr}
  \and
  Inria Nancy\\
  CAMUS Team\\
  615 Rue du Jardin-Botanique\\
  54600 Villers-lès-Nancy\\
  France\\
  ICube laboratory\\
  ICPS Team\\
  300 bd Sébastien Brant\\
  67412 Illkirch Cedex\\
  France\\
  \email{Berenger.Bramas@inria.fr}}

\begin{document}

\maketitle
\begin{abstract}
The sparse matrix/vector product (SpMV) is a fundamental operation in scientific computing. 
Having access to an efficient SpMV implementation is therefore critical, if not mandatory, to solve challenging numerical problems.
The ARM-based AFX64 CPU is a modern hardware component that equips one of the fastest supercomputers in the world. 
This CPU supports the Scalable Vector Extension (SVE) vectorization technology, which has been less investigated than the classic x86 instruction set architectures.
In this paper, we describe how we ported the SPC5 SpMV framework on AFX64 by converting AVX512 kernels to SVE.
In addition, we present performance results by comparing our kernels against a standard CSR kernel for both Intel-AVX512 and Fujitsu-ARM-SVE architectures.
\vspace{6pt}\textbf{Keywords:} SpMV, vectorization, AVX-512, SVE.
\end{abstract}

\section{Introduction}

The sparse matrix/vector product (SpMV) is a fundamental operation in scientific computing. 
It is the most important component of iterative linear solvers, which are widely used in finite element solvers. 
This is why SpMV has been and remains studied and improved.

Most of the studies work on the storage of sparse matrices, the implementation of SpMV kernels for novel hardware, or the combination of both.

In a previous work \cite{10.7717/peerj-cs.151}, we proposed a new sparse matrix storage format and its corresponding SpMV kernel in a framework called SPC5. 
The implementation was for x86 CPUs using the AVX512 instruction set architectures, and it was efficient for various types of data distribution.

In the current work, we are interested in porting this implementation on ARM SVE \cite{10.1109/MM.2017.35,ARMSVE,ARMSVE2} architecture. 
In other words, we aim at keeping the SPC5 storage format but create computational kernels that are efficient on ARM CPUs with SVE.

AVX512 and SVE instruction set architectures are different in their philosophies and features.
Consequently, as it is usually the case with vectorization, providing a new computational kernel is like solving a puzzle: we have the operation we want to perform on one side and the existing hardware instructions on the other side.

The contribution of the paper is to depict a new SpMV kernel for ARM SVE, and to demonstrate its performance on several sparse matrices of different shapes.
A secondary contribution is the description of our new AVX512 implementation, which is much simpler than the previous assembly implementation, while still delivering the same performance.

This paper is organized as follows.
In Section~\ref{sec:background}, we start by describing the vectorization principle, then the SpMV operation and the challenges of its efficient implementation, and finally provide the specificities of SPC5.
Then, in Section~\ref{sec:spc5}, we present our SPC5 implementation with SVE.
Finally, we study the performance of our implementation in Section~\ref{sec:study}.

\section{Background}
\label{sec:background}

\subsection{Vectorization}
\label{sec:vec}

\begin{figure}
    \centering
    \includegraphics[width=\textwidth, keepaspectratio]{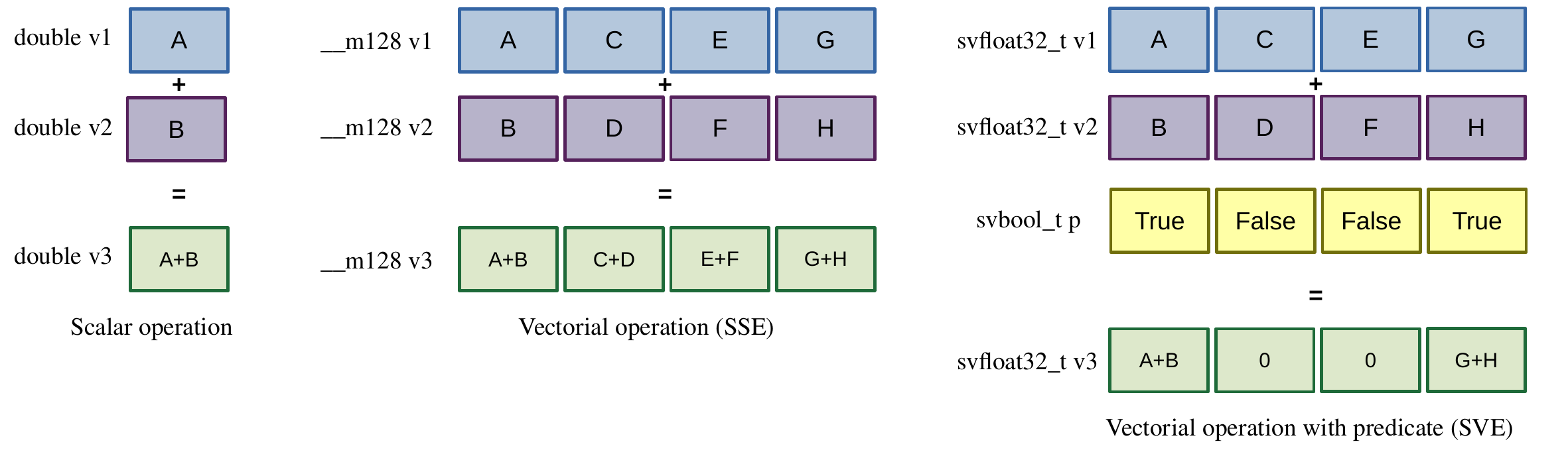}
    \caption{Illustration of a scalar operation, a vectorial operation and a vectorial operation with predicate.
             Each of the three instructions is performed with a single instruction.}
    \label{fig:vecop}
\end{figure}

Vectorization, also named SIMD for single instruction multiple data~\cite{flynn1966very}, is a key mechanism of modern processing units to increase the performance despite the clock frequency stagnation.
As its name suggests, the idea consists in working on several elements stored in vectors instead of scalar distinct elements.
As such, instead of performing operations on one element at a time, we perform the operations on vectors of elements using a vector instruction set architecture (ISA) that supports vector instructions.
We provide a schematic view of the concept in Figure~\ref{fig:vecop}.

Vectorization is straightforward when we aim to apply the same operation on all the elements of a vector.
However, the principle is challenging when we have divergence, i.e., we do not apply the exact same operations on all the elements, or when we need to perform data layout transformations, i.e., the input/output data blocks from the main memory that are loaded from (stored to) vectors are not contiguous, or we need to shuffle the data inside the vectors.

Moreover, not all instruction sets support the same operations, making each implementation specific to a given hardware.
Consequently, what could be done with a single instruction in a given instruction set, might need several instructions in another.
For example, non-contiguous stores (scatter), non-contiguous loads (gathers), or internal permutation/merging of vectors are not available in all existing instruction sets and not necessarily similar when they are supported.

Many computational algorithms use conditional statements, therefore several solutions have been proposed to manage vector divergences.
The first one is the single instruction multiple thread (SIMT) programming model, as used in CUDA and OpenCL.
While the programmer expresses its parallel algorithm as if independent execution threads would be used, it is actually large vector units that will perform the execution, where each thread will be an element of the vector.
The hardware takes care of the coherency during the execution.

The second mechanism is the use of a vector of predicates, where each predicate tells if an operation should be applied on an element of the vector.
When the elements of a vector should follow different execution paths (branches), all paths will be executed but predicate vectors will ensure to apply the correct operations.
The ARM SVE technology uses this mechanism, and most instructions can be used with a predicate vector.
Similar behavior can be obtained with classic x86 instruction sets using, for example, binary operations to merge several vectors obtained through different execution branches.

\subsection{Related Work on Vectorized with SVE}

Developing optimized kernels with SVE is a recent research topic \cite{8514923,8049003,aokioptimization,9513183,domke2021a64fx}.
A previous study \cite{alappat2021ecm} has focused on the modelling and tuning of the A64FX CPU. 
The authors implemented the SELL-C-$\sigma$ SpMV kernels and tuned it for this hardware. 
This kernel was originally made for GPUs but works well on CPUs too. However, the format is very different from the CSR format and requires a costly conversion step, which we aim to avoid. 
Additionally, the authors have performed important tuning for each matrix, by permuting the matrix or performing costly parameter optimization, where we want to provide a unique solution.

\subsection{SpMV}

The SpMV operation has been widely studied. 
This operation is memory bound in most cases with a low arithmetic intensity. 
Consequently, a naive vectorization usually does not provide significant benefits if the arithmetic intensity remains unchanged. 
This is why the storage of the sparse matrix is usually the central point of improvement.

Each new ISA can potentially help to create new storage formats that take less memory and/or that can be vectorized more efficiently.

For example, consider the more simple storage format called \emph{coordinates} (COO) or IJV, where each non-zero value (NNZ) is stored with a triple row index, column index and floating point value. 
In this case, for each NNZ we need two integers and one floating point value.
Not only this format is heavy but it is difficult to vectorized its corresponding SpMV kernel.

Another well-known storage format is the \emph{compressed sparse row} (CSR), where the values of the same row are stored contiguously such that there is no need to store an individual row index per value. 
With the CSR, each NNZ needs a single integer, which is the column index, decreasing the memory footprint up to 33\% compared to COO/IJV.

Following this idea, plenty of storage formats have been proposed. 
Many of them also tried to obtain a format that can be computed efficiently for a given architecture.

Some of the first block-based formats are the block compressed sparse row storage (BCSR) \cite{bib:tspsparse} and its extensions to larger blocks of variable dimension \cite{vuduc2003automatic,im2004sparsity} or to unaligned block compressed sparse row (UBCSR) \cite{bib:dssparse}. 
However, in these formats, the blocks have to be filled with zeros to be full.
For these formats, the blocks were aligned (the upper-left corner of the blocks start at a position multiple of the block size).
While the blocks are well suited for vectorization, the extra zeros can dramatically decrease the performance.

More recent work has focused on GPUs and manycore architectures.
Among them, the references are the ELLAPACK format \cite{liu2013efficient}, SELL-C-$\sigma$ \cite{kreutzer2014unified} defined as a variant of Sliced ELLPACK, and the CSR5 \cite{liu2015csr5} format that we used as reference in our previous study.

The Cuthill-McKee method from \cite{cuthill1969reducing} is a well-known technique for improving the bandwidth of a matrix to have good properties for LU decomposition. 
It does so by applying a breadth-first algorithm on a graph which represents the matrix structure. 
While the aim of this algorithm is not to improve the SpMV performance, the generated matrices may have better data locality.

Another method~\cite{bib:tspsparse} has been specifically designed to increase the number of contiguous values in rows and/or columns.
This method works by creating a graph from a matrix, where each column (or row) is a vertex and all the vertices are connected with weighted edges.
The weights represent the interest of putting two columns (or rows) contiguously.
By solving the traveling salesman problem (TSP) to obtain a path that goes through all the nodes but only once and that minimizes the total weight of the path, we can find a permutation of the sparse matrix that should be better divided into blocks.
This means that we should have fewer blocks and the blocks should contain more NNZ elements.

Several updates to the method have been presented in~\cite{bib:dssparse,pichel2005performance,bramas2016optimization} using different formulas. While the current study does not focus on the permutation of matrices, it is worth noting that enhancing the matrix's shape, as in other approaches, would likely lead to improved kernel efficiency by reducing the number of blocks.

\subsection{SPC5}
\label{sec:spc5}

The SPC5 format consists in using a block scheme without adding additional zeros.
SPC5 can be seen as an extension of the CSR format, but where the values of each row are split into blocks.
Each block starts with a NNZ at column \emph{c} and includes the next NNZ values until column \emph{c+VEC\_SIZE-1} if they exist.
Consequently, in the worst case a block contains a single value, and in the best case \emph{VEC\_SIZE} values.
Then, for each block, we use a mask of bits to indicate which of the NNZ values in the block exist.
As a result, in a poor configuration, SPC5 will have the same memory footprint as the CSR plus one bit mask per NNZ.
On the other hand, in the other extreme scenario, SPC5 will save one integer for each value added to a block since we can retrieve the corresponding column index from $c$ and the position of the NNZ in the block.

\begin{figure}
    \centering
    \includegraphics[width=\textwidth, keepaspectratio]{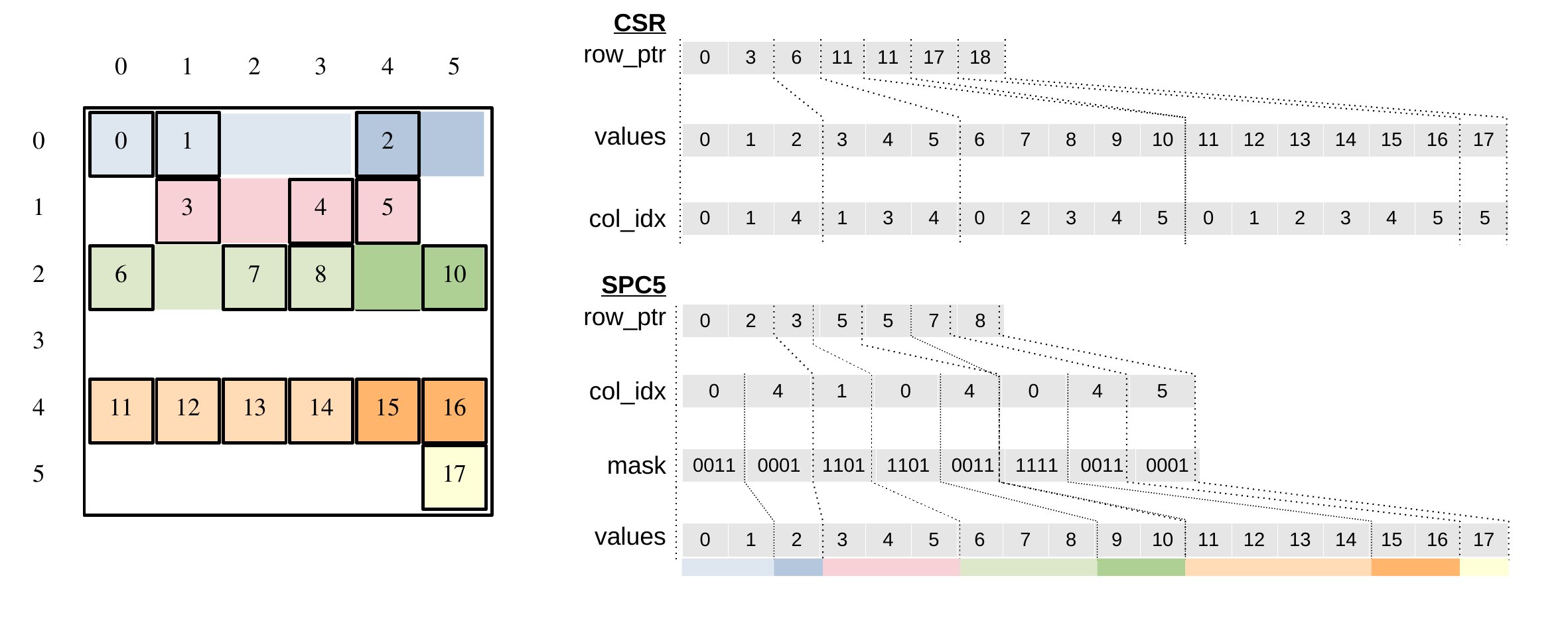}
    \caption{Illustration of the CSR and SPC5 formats.
             In this figure, we use the $\beta$(1,4) format, which means that each block is on a single row and of length 4.
            In the CSR format, the original \emph{row\_idx} array is compacted to have a single index per block instead of an index per NNZ.
            The mask array indicates the positions of the next NNZ in the block, and the corresponding column index can be obtained by summing the block's column index with the corresponding bit position in the mask.
            }
    \label{fig:spc5}
\end{figure}

The SPC5 format has been extended so that a block is mapped to several rows.
This is helpful if there are NNZ values closed (NNZ of consecutive rows that have closed column index) such that the values loaded from the vector \emph{x} can be used more than once and that the column index of the block is reused for more NNZ.

In the rest of the document, we refer to $\beta$(r, VEC\_SIZE) when the blocks are over \emph{r} rows and of length \emph{VEC\_SIZE}.
In the original study, we were also using blocks of $VEC\_SIZE/2$ but not in the current study.
We give an example of CSR and SPC5 in Figure~\ref{fig:spc5}.

\section{SPC5 Implementation}

We provide the SPC5 SpMV pseudocode in Algorithm~\ref{algo:spmvscalar}.

First, we initialize an index to progress in the array of NNZ values, at line~\ref{alg1:idxVal}.
Since we have no way to know where the values of a given block are located in the array, we have to increase the index with the number of values of each block that has been computed.
This is visible at line~\ref{alg1:idxValinc} for the scalar version, line~\ref{alg1:idxValincavx} for AVX512, and line~\ref{alg1:idxValincsve} for SVE.

At line~\ref{alg1:for_row}, we iterate over the rows with a step $r$. 
For each row segment, we iterate over the blocks at line~\ref{alg1:for_block}. 
For each block, we load its column index at line~\ref{alg1:store_idxCol}.

Then, we process each row of the block.
We start by getting the mask, at line~\ref{alg1:mask}, that tells us what are the NNZ that exist in the row of the block.
A naive implementation consists in testing the existence of each possible value, at line~\ref{alg1:testmaskscalar}, and performing the computation if needed, at line~\ref{alg1:testopcalar}. 
However, this loop over the NNZ can be done with a few instructions in AVX512, at line~\ref{alg1:avxop}. 
In this case, we load a vector from $x$ that matches the column index, and expand the NNZ from the value array into a vector.

\begin{figure}
    \centering
    \includegraphics[width=\textwidth, keepaspectratio]{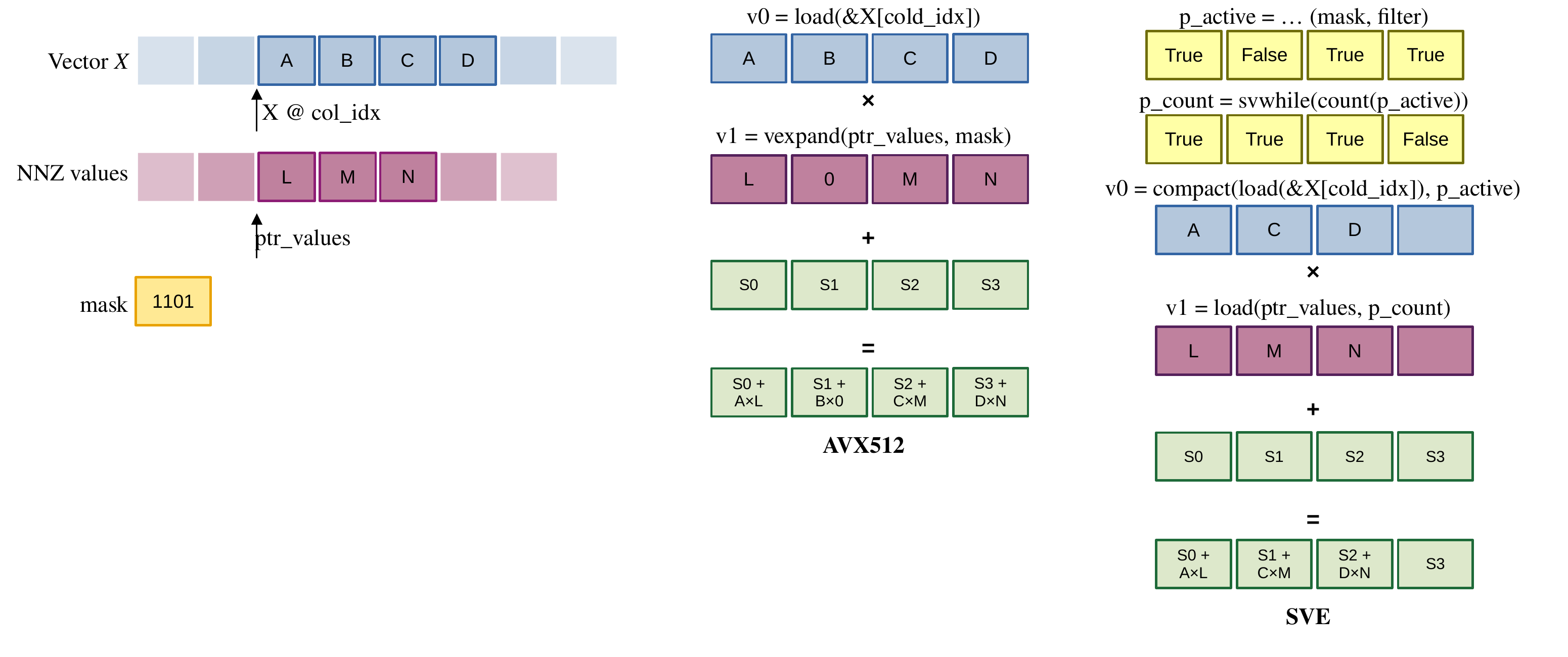}
    \caption{Illustration of the loading and computation of one row of a block.
            The mask is represented from the most significant bit (MSB) to the least significant bit (LSB), whereas the vector elements are represented from the first element to the last element.
            Hence, the $1$s in the mask $1101$ correspond to the elements N, M, and L (in this order).
            }
    \label{fig:expandorcompact}
\end{figure}

The expand operation does not exist in SVE. 
Consequently, the same behavior is obtained using different instructions. 
First, we need a filter vector that contains $2^i$ at position $i$, see line~\ref{alg1:filter}.
Then, we compute a binary \emph{and} operation between the filter vector and the mask, such that only the position for which a NNZ exist will not be zero.
We do this operation at line~\ref{alg1:filterop} and get the active elements at line~\ref{alg1:active}.
Second, instead of expanding the NNZ values, as with AVX512, we compact the values from $x$, at line~\ref{alg1:compact}.
Doing so, we can simply load the right number of NNZ and leave them contiguous in the resulting vector, before performing the computation.
A schematic view of the two approaches is provided in Figure~\ref{fig:expandorcompact}.

Finally, we update $y$ at the end of the algorithm (at line~\ref{alg1:updatey}) for each of the rows that have been proceeded.

\begin{algorithm2e}[h!]
\SetAlgoLined
\DontPrintSemicolon
\KwIn{x : vector to multiply with the matrix.
      mat : a matrix in the block format $\beta(r,c)$.
      r, c : the size of the blocks.}
\KwOut{y : the result of the product.}
 \Fn{spmv(x, mat, r, c, y)}
 {
        \tcp{Index to access the array's values}
        idxVal $\leftarrow$ 0\; \label{alg1:idxVal}
        \HiLi filter $\leftarrow$ [1 $<<$ 0, ...., 1 $<<$ VS-1]\; \label{alg1:filter}
        \For{idxRow $\leftarrow$ 0 \textbf{to} mat.numberOfRows-1 \textbf{inc by} $r$}{ \label{alg1:for_row}
            \HiLiScalar sum[r] $\leftarrow$ init\_scalar\_array(r, 0)\;
            \HiLi sum[r] $\leftarrow$ init\_simd\_array(r, 0)\;
            \For{idxBlock $\leftarrow$ mat.block\_rowptr[idxRow/r] \textbf{to} mat.block\_rowptr[idxRow/r+1]-1}{ \label{alg1:for_block}
                 idxCol $\leftarrow$ mat.block\_colidx[idxBlock]\; \label{alg1:store_idxCol}
                \For{idxRowBlock $\leftarrow$ 0 \textbf{to} r}{
                    valMask $\leftarrow$ mat.block\_masks[idxBlock $\times$ $r$ + idxRowBlock]\; \label{alg1:mask}
                    \tcp{The next loop can be vectorized with vexpand}
                    \HiLiScalar \For{k $\leftarrow$ 0 \textbf{to} $c$}{
                    \HiLiScalar    \If{bit\_shift(1 , k) BIT\_AND valMask}{ \label{alg1:testmaskscalar}
                    \HiLiScalar        sum[idxRowBlock] += x[idxCol+k] * mat.values[idxVal]\; \label{alg1:testopcalar}
                    \HiLiScalar       idxVal += 1\; \label{alg1:idxValinc}
                        }
                    }
                    \tcp{To replace the k-loop AVX512}
                    \HiLiAvx sum[idxRowBlock] += simd\_load(x[idxCol]) * \label{alg1:avxop}
                    \HiLiAvx       simd\_vexpand(mat.values[idxVal], valMask)\;
                    \HiLiAvx idxVal += popcount(valMask)\; \label{alg1:idxValincavx}
                    \tcp{To replace the k-loop SVE}
                    \HiLi mask\_vec = svand(svdup(valMask), filter)\;\label{alg1:filterop}
                    \HiLi active\_elts = svcmpne(mask\_vec, 0)\;\label{alg1:active}
                    \HiLi increment = count(active\_elts)\;
                    \HiLi xvals = svcompact(active\_elts, simd\_load(active\_elts, x[idxCol]))\;\label{alg1:compact}
                    \HiLi block = simd\_load(svwhile(0, increment), mat.values[idxVal])\;
                    \HiLi idxVal += increment\; \label{alg1:idxValincsve}
                    \HiLi sum[idxRowBlock] += block * xvals\;
                }
                
            }
            \For{idxRowBlock $\leftarrow$ idxRow \textbf{to} r+idxRow}{\label{alg1:updatey}
                \HiLiScalar y[idxRowBlock] += sum[r]\; 
                \HiLi y[idxRowBlock] += simd\_hsum(sum[r])\;\label{alg1:hsum}
            }
        }
 }
 \caption{SpMV for a matrix \emph{mat} in format $\beta(r,c)$.
          The lines in blue {\color{blue!90} \textbullet } are to compute in scalar and have to be replaced by the line in green {\color{green!90} \textbullet } to have the vectorized equivalent in SVE or in red {\color{red!90} \textbullet } with AVX.}
 \label{algo:spmvscalar}
\end{algorithm2e}

\subsection{Optimizing the Loading of $x$}
\label{sec:opt:x}

In AVX512, the values from $x$ are loaded into a SIMD vector without pruning.
This means that no matter how many NNZ are in the block or how many values we need from $x$, VEC\_SIZE values will be loaded from memory. 
It is possible to prune/filter the values, but this will imply an extra cost (i.e., using a more expensive instruction like gather) and would certainly have no benefit as the AVX512 SIMD load instruction is translated into an efficient memory transaction.
Consequently, in AVX512, the main optimization consists in loading the values from $x$ once for all the rows of a block, which allows accessing the memory once and using the resulting vector $r$ times.

\FloatBarrier

With SVE, it is different and we have mainly three alternatives:

\begin{itemize}
    \item Loading the values from $x$ without pruning, as with AVX512, and then compacting the obtained vector for each row of the block. We refer to this strategy as \emph{single $x$ load}.
    \item Loading a different vector for each row of the block, as shown in Algorithm~\ref{algo:spmvscalar}. 
    We refer to this strategy as \emph{partial $x$ load}.
    \item Combining the predicates of several rows of the block by merging their predicates/masks to load all the values that are needed by the block, but not more.
    In our study, we left this approach aside, as different tests we have conducted have shown poor performance.
\end{itemize}

The performance gains we can expect from the different strategies depend on the way the load is actually performed by the hardware.
In fact, the main point is to know if the hardware can make faster memory transactions when some elements of the predicate vector used in the load are false.
If the hardware actually loads VEC\_SIZE values from the memory but then only copies the ones that have their corresponding predicate value true to the SIMD vector, we should not expect any benefit.
Moreover, ARM SVE can be seen as an interface, hence it can be implemented by the hardware differently such that the behavior can also change from one vendor to another.

\subsection{Optimizing the Writing of the Result in $y$}
\label{sec:opt:y}

In the SIMD implementation, we have to perform the reduction (i.e., the horizontal sum) of $r$ vectors to add them to $y$ and store the result.

A straightforward approach is to call a single reduction instruction per vector, as both AVX512 and SVE support such an operation.
However, this means that we will perform $r$ individual summations between the reduced values and the values from $y$, and that we will also access the memory $r$ times (actually $2 \times r$, since we load values from $y$, perform the summation, and write back to $y$).

To avoid this, we propose a possible optimization that consists in performing the reduction of all the vectors manually to obtain a single vector as output, and then performing a vectorial summation with $y$.

With AVX512, this manual multi-reduction can be implemented by playing with AVX and SSE registers and using the horizontally add adjacent pairs instruction (\emph{hadd}).
The operation is done without any loop.

With SVE, we do this using odd/even interleave instructions (\emph{svuzp1} and \emph{svuzp2}). 
In this case, we need a loop because the length of the vectors is unknown at compile time.

\section{Performance Study}
\label{sec:study}
\subsection{Configuration}

We assess our method on two configurations:
\begin{itemize}
    \item Fujitsu-SVE: it is an ARMv8.2 A64FX - Fujitsu with 48 cores at 1.8 GHz and 512-bit SVE~\cite{a64fxmanual}, i.e. a vector can contain 16 single precision floating-point values or 8 double precision floating-point values.
    The node has 32 GB HBM2 memory arranged in four core memory groups (CMGs) with 12 cores and 8GB each, 64KB private L1 cache, and 8MB shared L2 cache per CMG.
    We use the GNU compiler 10.3.0.

    \item Intel-AVX512: it is a 2 $\times$ 18-core Cascade Lake Intel Xeon Gold 6240 at 2.6 GHz with AVX-512 (Advanced Vector 512-bit,  Foundation,  Conflict Detection, Byte and Word, Doubleword and Quadword Instructions, and Vector Length).
    The main memory consists of 190 GB DRAM memory arranged in two NUMA nodes. 
    Each CPU has 18 cores with 32KB private L1 cache, 1024KB private L2 cache, and 25MB shared L3 cache.
    We use the GNU compiler 11.2.0 and the MKL 2022.0.2.
\end{itemize}

\subsection{Test Cases}

We evaluated the performance of our proposed SPC5 SpMV kernels~\footnote{The code is available at \url{https://gitlab.inria.fr/bramas/spc5}} on a set of sparse matrices taken from the University of Florida sparse matrix collection (UF Collection) \cite{davis2011university}.
It includes a dense matrix of dimension 2048.
The results of the dense matrix are expected to provide an upper bound on the performance of the kernels, as all blocks will be full. 
Of course, our kernels are not well-designed or optimized for a dense matrix-vector product.
The properties of the matrices are given in Table~\ref{tab:matsetcomp}.

We evaluated the performance of four kernels: $\beta$(1, VS), $\beta$(2, VS), $\beta$(4, VS), and $\beta$(8, VS), where VS is the vector size.
We also provide the filling of the blocks when we format the matrices to the corresponding block sizes.
The filling can be up to 80\% for \emph{nd6k} but as low as 1\% for \emph{wikipedia-20060925}. (It is obviously 100\% for the dense matrix.)

We performed the computation in single precision (\emph{float/f32}) and double precision (\emph{double/f64}).

The original AVX implementation was written in assembly language, while our current implementation is written in C++ with intrinsic functions.
Consequently, the AVX and SVE kernels have very similar structures.

\input{tabmatrices}

\subsection{Results}

The results are organized as follows.
In the first part, we evaluate the difference of using the manual multi-reduction (described in Section~\ref{sec:opt:y}) vs. the native SIMD horizontal summation in Figures~\ref{res:perf:comp} (a) and~\ref{res:perf:comp} (b), for Fujitsu-SVE and Intel-AVX respectively.
For Fujitsu-SVE, we also evaluate the use of full vector load of $x$ (described in Section~\ref{sec:opt:y}).
Then, we provide the detailed results for all the matrices and the selected configuration in Figures~\ref{res:perf:full:sve} and~\ref{res:perf:full:avx}.
Finally, we provide an overview of the parallel performance when the computation is naively divided among the threads in Figure~\ref{res:perf:par}.


\input{figplotsve}
\input{figressve}

\paragraph{Comparisons of the Different Optimizations.}

We provide the performance results of the different implementations in Table~\ref{res:perf:comp}.

In Table~\ref{res:perf:comp} (a), we evaluate the use of manual multi-reduction and the single load of the $x$ vector for the Fujitsu-SVE architecture.
There is no difference in most cases, and it is always meaningless for $\beta$(2,VS).
The $\beta$(4,VS) and $\beta$(8,VS) kernels react differently with the optimizations and improve slightly.
The small change in using the multi-reduction comes from the small difference in latencies between the two approaches.
The \emph{reduce} instruction (\emph{addv}) has a latency of 12 cycles~\cite{a64fxmanual}.
Our multi-reduction has a latency of around 96 cycles for two vectors (considering the following latencies \emph{uzp1} $6$, \emph{uzp2} $6$, \emph{whilelt} $4$ and full \emph{vadd} $22$), and it is almost the same cost for 4 or 8 vectors.
Disabling the $x$ load optimization almost always degrades performance for the $\beta$(4,VS) kernel but seems to improve the performance for the $\beta$(8,VS).
This is surprising, as we would expect that the larger the blocks would be, the more benefit we would have to load the vector from $x$ completely.
From our understanding, the cost of a load depends on the location of the data it requests but not on the fact that the data it requests could be located in different cache lines.
This explains why the optimization has a limited impact, as a partial load will move the data to the cache and speedup the next partial loads.
Since the $\beta$(4,VS) is faster than $\beta$(8,VS), we consider the best configuration to be with both optimizations turned on.

In Table~\ref{res:perf:comp} (b), we evaluate the use of manual multi-reduction on Intel-AVX512 architecture.
The performance increases slightly with the use of manual multi-reduction in some cases.
For instance, the best performance on average is obtained with $\beta$(4,VS) and for this configuration, using the manual multi-reduction has no impact (double) or increases the speedup by $0.1$ (float).
The explanation is as follows: the \emph{reduce} intrinsic function (\emph{\_mm512\_reduce\_add\_ps}) is not actually a real hardware instruction, but a call to a function provided by the compiler~\cite{varela2022manycore}.
Its implementation~\footnote{\href{https://github.com/gcc-mirror/gcc/blob/9d7e19255c06e05ad791e9bf5aefc4783a12c4f9/gcc/config/i386/avx512fintrin.h\#L15928}{https://github.com/gcc-mirror/gcc/blob/9d7e19255c06e05ad791e9bf5aefc4783a12c4f9}
\href{https://github.com/gcc-mirror/gcc/blob/9d7e19255c06e05ad791e9bf5aefc4783a12c4f9/gcc/config/i386/avx512fintrin.h\#L15928}{/gcc/config/i386/avx512fintrin.h\#L15928}} is very similar to our manual multi-reduction, with the main difference being that we try to factorize the instructions to reduce several SIMD vectors at the same time. 
This allows us to obtain a SIMD vector as output, which can then be used to update $y$ with vectorized instructions.
In the end, the performance difference between the two approaches is limited. 
However, for the rest of the study, we consider that the best Intel-AVX512 configuration is to use manual multi-reduction.

\begin{landscape}

\input{figcomparmavxtab}
\end{landscape}

\paragraph{Best Configuration Detailed Results.}

We provide the complete results for Fujitsu-SVE in Figures~\ref{res:perf:full:sve} and~\ref{res:plotsve}. 
The results for Intel-AVX are shown in Figures~\ref{res:perf:full:avx} and~\ref{res:plotavx}.

In Figure~\ref{res:perf:full:sve}, we can see that the performance of the SPC5 kernels is clearly related to the block filling. 
The performance model can be described as a constant cost per block that seems independent of the number of blocks or the number of NNZ. 
This means that the performance can be easily predicted from the block filling.

We also note that the performance increases as we increase the size of the blocks up to 4×VS, but then it decreases for $8 \times VS$. 
This is more visible in Figure~\ref{res:perf:full:sve}, where $\beta$(8, VS) is the slowest SPC5 kernel in most cases.

The behaviors in single and double precision are similar. 
For some matrices, such as ns3Da, SPC5 is even slower than a simple CSR implementation. 
This means that the overhead of using vectorial instructions outweighs the benefits of vectorization since the block filling is very low.

The computation on the matrix TSOPF, which has a very high block filling, achieves performance almost equivalent to the dense matrix case. 
Finally, we can see the average performance in Figure~\ref{res:perf:full:sve} (last bars). 
While the speedup against CSR is significant, the raw performance is low compared to the peak performance of the machine.

The results for Intel-AVX are slightly different. 
In Figure~\ref{res:perf:full:avx}, we can see that while there is a correlation between the block filling and the performance, the relationship is less clear than for Fujitsu-SVE.

We also note that the performance increases with the block size, such that the best performance is achieved with $\beta$(8, VS). 
This is even more visible in Figure~\ref{res:perf:full:avx}.

Contrary to Fujitsu-SVE, the performance obtained for TSOPF is not close to the dense matrix case. 
This means that while the blocks are almost full, the fact that we have to jump over the vector $x$ has a negative impact on the performance.

The performance of SPC5 on Intel-AVX is higher than those obtained with Fujitsu-SVE for almost all matrices. 
Finally, SPC5 is faster than the Intel MKL CSR kernel for most matrices, but can be slower if there are less than two values per block.

\input{figplotavx}

\input{figresavx}
\paragraph{Parallel Performance Overview.}

In Figure~\ref{res:perf:par}, we provide the results for the parallel executions.
For the Fujitsu-SVE hardware, Figure~\ref{res:perf:par:sve}, for some matrices the speedup is above 42 (the number of cores).
This is possible because the matrices are split and allocated by the threads such that each thread has its data on the memory nodes that correspond to its CPU core.
In addition, the split of the matrices and the use of all the cores can result in using the cache more efficiently.

For the Intel-AVX512 hardware, Figure~\ref{res:perf:par:avx}, the executions on the dense matrix have poor performance for small blocks.
This is clear that the $x$ vector will be fully loaded for each row, such that the cache performance is tied to the final execution performance.
We can notice that the speedup around 15 is far from the number of CPU cores (36).
The workload balance between the threads is similar to the Fujitsu-SVE configuration, therefore, we consider that the difference comes from the memory organization and use.

\input{figrespar}

\section{Conclusion}

We have presented a new version of our SPC5 framework, which remains efficient on architectures with AVX512 and is now compatible with ARM architectures with SVE.
The same sparse matrix format can be used to target both ISA, allowing for interoperability and the use of a single framework on x86 and ARM-based architectures.

The SPC5's SpMV kernels are implemented differently, as they rely on an expand mechanism of the NNZ (AVX512) or a compaction of the $x$ vector (SVE).
The performances we obtained are usually higher than a simple CSR kernel if there is more than a single NNZ per block.
The $\beta$(1,$*$) format has a low conversion cost as it leaves the array of NNZ unchanged compared to CSR, which makes it easy to plug in existing CSR-based applications.

In a future work, we would like to investigate if we could use a hybrid format, i.e., a format where we could have blocks of different sizes including blocks of scalar, to avoid using vectorial instructions when there is no benefit.

\section{Acknowledgement}
This work used the Isambard 2 UK National Tier-2 HPC Service~\footnote{\url{http://gw4.ac.uk/isambard/}} operated by GW4 and the UK Met Office, which is an EPSRC project (EP/T022078/1).
We also used the PlaFRIM experimental testbed, supported by Inria, CNRS (LABRI and IMB), Universite de Bordeaux, Bordeaux INP and Conseil Regional d’Aquitaine~\footnote{\url{https://www.plafrim.fr}}. 
In addition, this work used the Farm-SVE library~\cite{bramas:hal-02906179v1}.

\bibliographystyle{splncs03}
\bibliography{references}


%
%


\end{document}

%% file: tabmatrices.tex
\begin{table}[ht]
{\footnotesize
\begin{tabular}{|l||c||c||@{}c@{}|@{}c@{}|@{}c@{}|@{}c@{}|@{}c@{}|}
\hline
\rule{0pt}{4ex} Name & Dim & NNZ & $\frac{NNZ}{N_{rows}}$ & $\beta(1,VS)$ & $\beta(2,VS)$ & $\beta(4,VS)$ & $\beta(8,VS)$  \\
\hline \hline
bundle & 513351 & 20208051 & 39.365 & 72\% $|$ 55\%  & 70\% $|$ 54\%  & 64\% $|$ 50\%  & 51\% $|$ 46\%  \\
CO & 221119 & 7666057 & 34.6694 & 18\% $|$ 9\%  & 18\% $|$ 9\%  & 17\% $|$ 9\%  & 16\% $|$ 8\%  \\
crankseg & 63838 & 14148858 & 221.637 & 66\% $|$ 49\%  & 59\% $|$ 44\%  & 49\% $|$ 37\%  & 38\% $|$ 29\%  \\
dense & 2048 & 4194304 & 2048 & 100\% $|$ 100\%  & 100\% $|$ 100\%  & 100\% $|$ 100\%  & 100\% $|$ 100\%  \\
dielFilterV2real & 1157456 & 48538952 & 41.9359 & 31\% $|$ 20\%  & 22\% $|$ 14\%  & 15\% $|$ 10\%  & 11\% $|$ 7\%  \\
Emilia & 923136 & 41005206 & 44.4195 & 50\% $|$ 31\%  & 43\% $|$ 28\%  & 34\% $|$ 24\%  & 24\% $|$ 18\%  \\
FullChip & 2987012 & 26621990 & 8.91258 & 24\% $|$ 13\%  & 17\% $|$ 10\%  & 13\% $|$ 7\%  & 8\% $|$ 5\%  \\
Hook & 1498023 & 60917445 & 40.6652 & 51\% $|$ 34\%  & 43\% $|$ 29\%  & 33\% $|$ 23\%  & 24\% $|$ 17\%  \\
in-2004 & 1382908 & 16917053 & 12.233 & 48\% $|$ 31\%  & 38\% $|$ 25\%  & 30\% $|$ 19\%  & 21\% $|$ 14\%  \\
ldoor & 952203 & 46522475 & 48.8577 & 87\% $|$ 55\%  & 79\% $|$ 51\%  & 67\% $|$ 44\%  & 51\% $|$ 34\%  \\
mixtank & 29957 & 1995041 & 66.5968 & 31\% $|$ 20\%  & 24\% $|$ 16\%  & 17\% $|$ 11\%  & 12\% $|$ 8\%  \\
nd6k & 18000 & 6897316 & 383.184 & 80\% $|$ 71\%  & 76\% $|$ 68\%  & 71\% $|$ 64\%  & 64\% $|$ 58\%  \\
ns3Da & 20414 & 1679599 & 82.2768 & 14\% $|$ 7\%  & 8\% $|$ 4\%  & 4\% $|$ 2\%  & 2\% $|$ 1\%  \\
pdb1HYS & 36417 & 4344765 & 119.306 & 77\% $|$ 65\%  & 72\% $|$ 60\%  & 63\% $|$ 54\%  & 54\% $|$ 46\%  \\
pwtk & 217918 & 11634424 & 53.389 & 74\% $|$ 56\%  & 74\% $|$ 55\%  & 73\% $|$ 54\%  & 65\% $|$ 53\%  \\
RM07R & 381689 & 37464962 & 98.1557 & 61\% $|$ 41\%  & 51\% $|$ 34\%  & 40\% $|$ 28\%  & 31\% $|$ 25\%  \\
Serena & 1391349 & 64531701 & 46.3807 & 51\% $|$ 34\%  & 43\% $|$ 29\%  & 33\% $|$ 23\%  & 24\% $|$ 17\%  \\
Si41Ge41H72 & 185639 & 15011265 & 80.8627 & 32\% $|$ 18\%  & 31\% $|$ 17\%  & 28\% $|$ 15\%  & 22\% $|$ 13\%  \\
Si87H76 & 240369 & 10661631 & 44.3553 & 21\% $|$ 11\%  & 21\% $|$ 11\%  & 20\% $|$ 10\%  & 17\% $|$ 9\%  \\
spal & 10203 & 46168124 & 4524.96 & 74\% $|$ 69\%  & 45\% $|$ 37\%  & 25\% $|$ 23\%  & 13\% $|$ 12\%  \\
torso1 & 116158 & 8516500 & 73.3182 & 81\% $|$ 63\%  & 80\% $|$ 62\%  & 77\% $|$ 59\%  & 58\% $|$ 55\%  \\
TSOPF & 38120 & 16171169 & 424.217 & 94\% $|$ 88\%  & 93\% $|$ 87\%  & 92\% $|$ 85\%  & 89\% $|$ 82\%  \\
wikipedia-20060925 & 2983494 & 37269096 & 12.4918 & 13\% $|$ 6\%  & 6\% $|$ 3\%  & 3\% $|$ 1\%  & 1\% $|$ 0\%  \\
\hline
\end{tabular}
}
\caption{Matrix set for computation and performance analysis.
         We provide the percentage of filling of the blocks for double (left) and single (right) precision.}
\label{tab:matsetcomp}
\end{table}

%% file: figplotsve.tex
\begin{figure}[h!]
\centering
    \begin{subfigure}[c]{\textwidth}
    \begin{tikzpicture}
    \begin{axis}[
            height=.25\textheight,
            width=\textwidth,
            ymin=0,
            xmin=0,
            clip=false,
            axis lines*=left,
            legend style={at={(0.9,1.3)}, anchor=east,legend columns=4, font=\footnotesize},
            ylabel={GFlop/s},
            xlabel={Average NNZ per blocks},
            yticklabel style = {font=\footnotesize,xshift=0.5ex},
            xticklabel style = {font=\footnotesize,yshift=0.5ex},
            ylabel style = {font=\footnotesize},
            xlabel style = {font=\footnotesize},
    every node near coord/.style={font=\tiny}
        ]
\addplot[draw=color_1, fill=color_1!30!white, only marks] table[x index=8, y index=9] {data.txt};
\addplot[draw=color_2, fill=color_2!30!white, only marks] table[x index=10, y index=11] {data.txt};
\addplot[draw=color_3, fill=color_3!30!white, only marks] table[x index=12, y index=13] {data.txt};
\addplot[draw=color_4, fill=color_4!30!white, only marks] table[x index=14, y index=15] {data.txt};

 \legend{SPC5 $\beta$(1$\cdot$VS), SPC5 $\beta$(2$\cdot$VS), SPC5 $\beta$(4$\cdot$VS), SPC5 $\beta$(8$\cdot$VS)}
        \end{axis}
        \end{tikzpicture}
        \caption{Double precision (f64).}
        \label{res:plotsve:f64}
    \end{subfigure}
    \hfill
    \vspace{-3mm}
    \begin{subfigure}[c]{\textwidth}
    \begin{tikzpicture}
    \begin{axis}[
            height=.25\textheight,
            width=\textwidth,
            ymin=0,
            xmin=0,
            clip=false,
            axis lines*=left,
            legend style={at={(0.9,1.3)}, anchor=east,legend columns=4, font=\footnotesize},
            ylabel={GFlop/s},
            xlabel={Average NNZ per blocks},
            yticklabel style = {font=\footnotesize,xshift=0.5ex},
            xticklabel style = {font=\footnotesize,yshift=0.5ex},
            ylabel style = {font=\footnotesize},
            xlabel style = {font=\footnotesize},
    every node near coord/.style={font=\tiny}
        ]
\addplot[draw=color_1, fill=color_1!30!white, only marks] table[x index=0, y index=1] {data.txt};
\addplot[draw=color_2, fill=color_2!30!white, only marks] table[x index=2, y index=3] {data.txt};
\addplot[draw=color_3, fill=color_3!30!white, only marks] table[x index=4, y index=5] {data.txt};
\addplot[draw=color_4, fill=color_4!30!white, only marks] table[x index=6, y index=7] {data.txt};

        \end{axis}
        \end{tikzpicture}
        \caption{Single precision (f32).}
        \label{res:plotsve:f32}
    \end{subfigure}
        \caption{Performance in Giga Flop per second for sequential computation in double and single precision for our SPC5 kernels on Fujitsu-SVE architecture for all the matrices of the test set.
                 }
        \label{res:plotsve}
\end{figure}
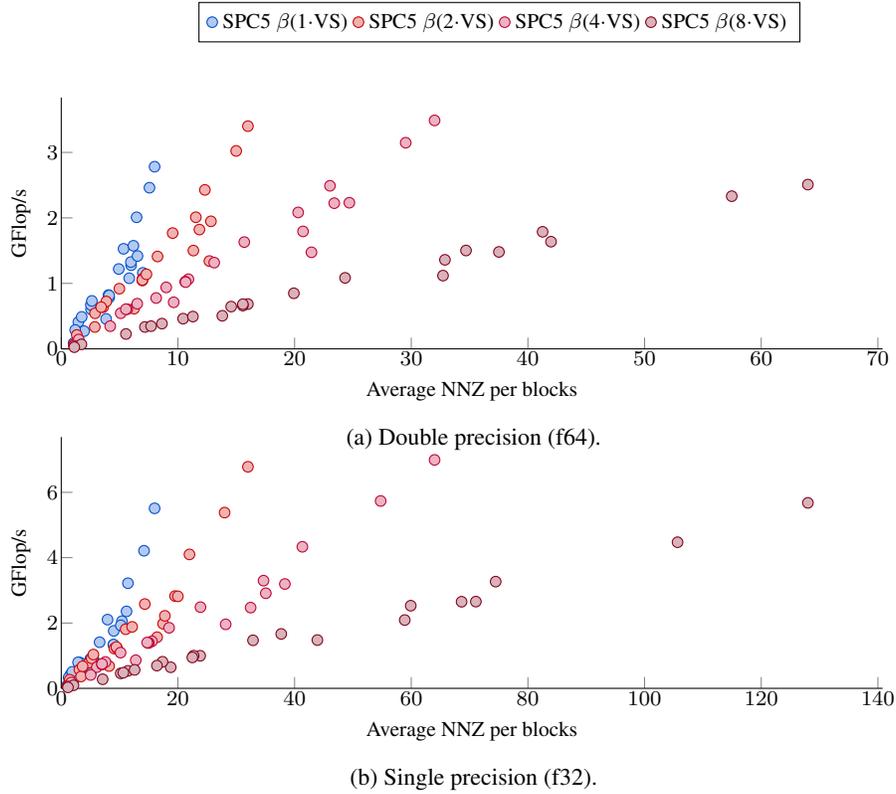

%% file: figressve.tex
\begin{figure}[h!]
\centering
    \begin{subfigure}[c]{\textwidth}
        \begin{tikzpicture}
        \begin{axis}[
                height=.18\textheight,
                width=\textwidth,
                ybar=0pt,
                ymax=7,
                ymin=0,
                clip=false,
                axis lines*=left,
                bar width=3.9pt,
                enlarge x limits=0.05,
                legend style={at={(0.95,1.7)}, anchor=east,legend columns=4, font=\footnotesize},
                ylabel={GFlop/s},
                symbolic x coords={bundle (f64),bundle (f32),CO (f64),CO (f32),crankseg (f64),crankseg (f32),dense (f64),dense (f32),dielFilterV2real (f64),dielFilterV2real (f32),Emilia (f64),Emilia (f32)},
                xticklabels={bundle (f64),bundle (f32),CO (f64),CO (f32),crankseg (f64),crankseg (f32),dense (f64),dense (f32),dielFilterV2real (f64),dielFilterV2real (f32),Emilia (f64),Emilia (f32)},
                xtick=data,
                nodes near coords,
                point meta=explicit symbolic,
                every node near coord/.append style={font=\footnotesize},
                nodes near coords align={vertical},
                yticklabel style = {font=\footnotesize,xshift=0.5ex},
                xticklabel style = {font=\footnotesize,yshift=0.5ex},
                ylabel style = {font=\footnotesize},
                xlabel style = {font=\footnotesize},
                every axis legend/.append style={nodes={right}},
                every node near coord/.append style={font=\tiny, rotate=90, anchor=west},
                x tick label style={rotate=50,anchor=east},
                ymajorgrids = true,
            ]
\addplot[draw=color_0!65!black, fill=color_0!30!white] coordinates {  (bundle (f64), 0.383652)  (bundle (f32), 0.386417)  (CO (f64), 0.348644)  (CO (f32), 0.354306)  (crankseg (f64), 0.383573)  (crankseg (f32), 0.387428)  (dense (f64), 0.394231)  (dense (f32), 0.395927)  (dielFilterV2real (f64), 0.357005)  (dielFilterV2real (f32), 0.361176)  (Emilia (f64), 0.359177)  (Emilia (f32), 0.362692)  };
\addplot[draw=color_1!65!black, fill=color_1!30!white] coordinates {  (bundle (f64), 1.07839)[2.8]  (bundle (f32), 1.33246)[3.4]  (CO (f64), 0.409804)[1.2]  (CO (f32), 0.422459)[1.2]  (crankseg (f64), 1.52758)[4.0]  (crankseg (f32), 2.10396)[5.4]  (dense (f64), 2.78158)[7.1]  (dense (f32), 5.5103)[13.9]  (dielFilterV2real (f64), 0.604207)[1.7]  (dielFilterV2real (f32), 0.676784)[1.9]  (Emilia (f64), 0.821739)[2.3]  (Emilia (f32), 0.897862)[2.5]  };
\addplot[draw=color_2!65!black, fill=color_2!30!white] coordinates {  (bundle (f64), 1.50107)[3.9]  (bundle (f32), 1.97827)[5.1]  (CO (f64), 0.543714)[1.6]  (CO (f32), 0.571626)[1.6]  (crankseg (f64), 1.76802)[4.6]  (crankseg (f32), 2.58005)[6.7]  (dense (f64), 3.40071)[8.6]  (dense (f32), 6.78269)[17.1]  (dielFilterV2real (f64), 0.636439)[1.8]  (dielFilterV2real (f32), 0.785992)[2.2]  (Emilia (f64), 1.07171)[3.0]  (Emilia (f32), 1.21758)[3.4]  };
\addplot[draw=color_3!65!black, fill=color_3!30!white] coordinates {  (bundle (f64), 1.79355)[4.7]  (bundle (f32), 2.4712)[6.4]  (CO (f64), 0.605473)[1.7]  (CO (f32), 0.6498)[1.8]  (crankseg (f64), 1.63068)[4.3]  (crankseg (f32), 2.4854)[6.4]  (dense (f64), 3.48845)[8.8]  (dense (f32), 6.99344)[17.7]  (dielFilterV2real (f64), 0.544181)[1.5]  (dielFilterV2real (f32), 0.729241)[2.0]  (Emilia (f64), 1.06338)[3.0]  (Emilia (f32), 1.43999)[4.0]  };
\addplot[draw=color_4!65!black, fill=color_4!30!white] coordinates {  (bundle (f64), 1.35986)[3.5]  (bundle (f32), 2.09062)[5.4]  (CO (f64), 0.461201)[1.3]  (CO (f32), 0.534849)[1.5]  (crankseg (f64), 1.08402)[2.8]  (crankseg (f32), 1.66578)[4.3]  (dense (f64), 2.5084)[6.4]  (dense (f32), 5.67819)[14.3]  (dielFilterV2real (f64), 0.334245)[0.9]  (dielFilterV2real (f32), 0.458877)[1.3]  (Emilia (f64), 0.684056)[1.9]  (Emilia (f32), 0.996511)[2.7]  };

 \legend{CSR, SPC5 $\beta$(1$\cdot$VS), SPC5 $\beta$(2 $\cdot$ VS), SPC5 $\beta$(4$\cdot$VS), SPC5 $\beta$(8$\cdot$VS)}
        \end{axis}
        \end{tikzpicture}
    \end{subfigure}
    \hfill
    \vspace{-3mm}
    \begin{subfigure}[c]{\textwidth}
        \begin{tikzpicture}
        \begin{axis}[
                height=.18\textheight,
                width=\textwidth,
                ybar=0pt,
                ymax=7,
                ymin=0,
                clip=false,
                axis lines*=left,
                bar width=3.9pt,
                enlarge x limits=0.05,
                legend style={at={(0.95,1.3)}, anchor=east,legend columns=6, font=\footnotesize},
                ylabel={GFlop/s},
                symbolic x coords={FullChip (f64),FullChip (f32),Hook (f64),Hook (f32),in-2004 (f64),in-2004 (f32),ldoor (f64),ldoor (f32),mixtank (f64),mixtank (f32),nd6k (f64),nd6k (f32)},
                xticklabels={FullChip (f64),FullChip (f32),Hook (f64),Hook (f32),in-2004 (f64),in-2004 (f32),ldoor (f64),ldoor (f32),mixtank (f64),mixtank (f32),nd6k (f64),nd6k (f32)},
                xtick=data,
                nodes near coords,
                point meta=explicit symbolic,
                every node near coord/.append style={font=\footnotesize},
                yticklabel style = {font=\footnotesize,xshift=0.5ex},
                xticklabel style = {font=\footnotesize,yshift=0.5ex},
                ylabel style = {font=\footnotesize},
                xlabel style = {font=\footnotesize},
                every axis legend/.append style={nodes={right}},
                every node near coord/.append style={font=\tiny, rotate=90, anchor=west},
                x tick label style={rotate=50,anchor=east},
                ymajorgrids = true,
            ]

\addplot[draw=color_0!65!black, fill=color_0!30!white] coordinates {  (FullChip (f64), 0.299325)  (FullChip (f32), 0.31086)  (Hook (f64), 0.357144)  (Hook (f32), 0.361609)  (in-2004 (f64), 0.335278)  (in-2004 (f32), 0.340649)  (ldoor (f64), 0.350317)  (ldoor (f32), 0.357269)  (mixtank (f64), 0.370127)  (mixtank (f32), 0.373368)  (nd6k (f64), 0.388165)  (nd6k (f32), 0.391935)  };
\addplot[draw=color_1!65!black, fill=color_1!30!white] coordinates {  (FullChip (f64), 0.270498)[0.9]  (FullChip (f32), 0.270806)[0.9]  (Hook (f64), 0.783076)[2.2]  (Hook (f32), 0.891661)[2.5]  (in-2004 (f64), 0.457029)[1.4]  (in-2004 (f32), 0.465207)[1.4]  (ldoor (f64), 1.16058)[3.3]  (ldoor (f32), 1.34051)[3.8]  (mixtank (f64), 0.672432)[1.8]  (mixtank (f32), 0.787317)[2.1]  (nd6k (f64), 2.00921)[5.2]  (nd6k (f32), 3.21793)[8.2]  };
\addplot[draw=color_2!65!black, fill=color_2!30!white] coordinates {  (FullChip (f64), 0.334224)[1.1]  (FullChip (f32), 0.36314)[1.2]  (Hook (f64), 1.04428)[2.9]  (Hook (f32), 1.23442)[3.4]  (in-2004 (f64), 0.613565)[1.8]  (in-2004 (f32), 0.682525)[2.0]  (ldoor (f64), 1.34105)[3.8]  (ldoor (f32), 1.56525)[4.4]  (mixtank (f64), 0.723486)[2.0]  (mixtank (f32), 0.920195)[2.5]  (nd6k (f64), 2.42754)[6.3]  (nd6k (f32), 4.09958)[10.5]  };
\addplot[draw=color_3!65!black, fill=color_3!30!white] coordinates {  (FullChip (f64), 0.346735)[1.2]  (FullChip (f32), 0.409556)[1.3]  (Hook (f64), 1.0261)[2.9]  (Hook (f32), 1.39385)[3.9]  (in-2004 (f64), 0.711475)[2.1]  (in-2004 (f32), 0.860654)[2.5]  (ldoor (f64), 1.47457)[4.2]  (ldoor (f32), 1.95946)[5.5]  (mixtank (f64), 0.605331)[1.6]  (mixtank (f32), 0.807753)[2.2]  (nd6k (f64), 2.49024)[6.4]  (nd6k (f32), 4.33359)[11.1]  };
\addplot[draw=color_4!65!black, fill=color_4!30!white] coordinates {  (FullChip (f64), 0.227531)[0.8]  (FullChip (f32), 0.277981)[0.9]  (Hook (f64), 0.66088)[1.9]  (Hook (f32), 0.999811)[2.8]  (in-2004 (f64), 0.505279)[1.5]  (in-2004 (f32), 0.645038)[1.9]  (ldoor (f64), 1.11881)[3.2]  (ldoor (f32), 1.47818)[4.1]  (mixtank (f64), 0.346937)[0.9]  (mixtank (f32), 0.4785)[1.3]  (nd6k (f64), 1.78749)[4.6]  (nd6k (f32), 3.2668)[8.3]  };
        \end{axis}
        \end{tikzpicture}
    \end{subfigure}
    \hfill
    \vspace{-3mm}
\begin{subfigure}[c]{\textwidth}
        \begin{tikzpicture}
        \begin{axis}[
                height=.18\textheight,
                width=\textwidth,
                ybar=0pt,
                ymax=7,
                ymin=0,
                clip=false,
                axis lines*=left,
                bar width=3.9pt,
                enlarge x limits=0.05,
                legend style={at={(0.55,1.3)}, anchor=east,legend columns=2, font=\footnotesize},
                ylabel={GFlop/s},
                xlabel={Matrices},
                symbolic x coords={ns3Da (f64),ns3Da (f32),pdb1HYS (f64),pdb1HYS (f32),pwtk (f64),pwtk (f32),RM07R (f64),RM07R (f32),Serena (f64),Serena (f32),Si41Ge41H72 (f64),Si41Ge41H72 (f32)},
                xticklabels={ns3Da (f64),ns3Da (f32),pdb1HYS (f64),pdb1HYS (f32),pwtk (f64),pwtk (f32),RM07R (f64),RM07R (f32),Serena (f64),Serena (f32),Si41Ge41H72 (f64),Si41Ge41H72 (f32)},
                xtick=data,
                nodes near coords,
                point meta=explicit symbolic,
                every node near coord/.append style={font=\footnotesize},
                yticklabel style = {font=\footnotesize,xshift=0.5ex},
                xticklabel style = {font=\footnotesize,yshift=0.5ex},
                ylabel style = {font=\footnotesize},
                xlabel style = {font=\footnotesize},
                every axis legend/.append style={nodes={right}},
                every node near coord/.append style={font=\tiny, rotate=90, anchor=west},
                x tick label style={rotate=50,anchor=east},
                ymajorgrids = true,
            ]

\addplot[draw=color_0!65!black, fill=color_0!30!white] coordinates {  (ns3Da (f64), 0.363055)  (ns3Da (f32), 0.371543)  (pdb1HYS (f64), 0.381312)  (pdb1HYS (f32), 0.383346)  (pwtk (f64), 0.3668)  (pwtk (f32), 0.369147)  (RM07R (f64), 0.37593)  (RM07R (f32), 0.380277)  (Serena (f64), 0.358909)  (Serena (f32), 0.363688)  (Si41Ge41H72 (f64), 0.37103)  (Si41Ge41H72 (f32), 0.3759)  };
\addplot[draw=color_1!65!black, fill=color_1!30!white] coordinates {  (ns3Da (f64), 0.288488)[0.8]  (ns3Da (f32), 0.328004)[0.9]  (pdb1HYS (f64), 1.57372)[4.1]  (pdb1HYS (f32), 2.05671)[5.4]  (pwtk (f64), 1.27728)[3.5]  (pwtk (f32), 1.7585)[4.8]  (RM07R (f64), 1.21943)[3.2]  (RM07R (f32), 1.41237)[3.7]  (Serena (f64), 0.814908)[2.3]  (Serena (f32), 0.926976)[2.5]  (Si41Ge41H72 (f64), 0.732548)[2.0]  (Si41Ge41H72 (f32), 0.796106)[2.1]  };
\addplot[draw=color_2!65!black, fill=color_2!30!white] coordinates {  (ns3Da (f64), 0.211759)[0.6]  (ns3Da (f32), 0.264555)[0.7]  (pdb1HYS (f64), 2.00984)[5.3]  (pdb1HYS (f32), 2.82405)[7.4]  (pwtk (f64), 1.82164)[5.0]  (pwtk (f32), 2.21951)[6.0]  (RM07R (f64), 1.4106)[3.8]  (RM07R (f32), 1.81001)[4.8]  (Serena (f64), 1.05279)[2.9]  (Serena (f32), 1.26486)[3.5]  (Si41Ge41H72 (f64), 0.918831)[2.5]  (Si41Ge41H72 (f32), 1.03413)[2.8]  };
\addplot[draw=color_3!65!black, fill=color_3!30!white] coordinates {  (ns3Da (f64), 0.142936)[0.4]  (ns3Da (f32), 0.179414)[0.5]  (pdb1HYS (f64), 2.08341)[5.5]  (pdb1HYS (f32), 3.29653)[8.6]  (pwtk (f64), 2.22461)[6.1]  (pwtk (f32), 2.91072)[7.9]  (RM07R (f64), 1.31686)[3.5]  (RM07R (f32), 1.85414)[4.9]  (Serena (f64), 1.01935)[2.8]  (Serena (f32), 1.38648)[3.8]  (Si41Ge41H72 (f64), 0.939188)[2.5]  (Si41Ge41H72 (f32), 1.09308)[2.9]  };
\addplot[draw=color_4!65!black, fill=color_4!30!white] coordinates {  (ns3Da (f64), 0.0691491)[0.2]  (ns3Da (f32), 0.0936001)[0.3]  (pdb1HYS (f64), 1.50229)[3.9]  (pdb1HYS (f32), 2.52969)[6.6]  (pwtk (f64), 1.63692)[4.5]  (pwtk (f32), 2.65109)[7.2]  (RM07R (f64), 0.84881)[2.3]  (RM07R (f32), 1.47013)[3.9]  (Serena (f64), 0.682398)[1.9]  (Serena (f32), 0.946922)[2.6]  (Si41Ge41H72 (f64), 0.645536)[1.7]  (Si41Ge41H72 (f32), 0.813349)[2.2]  };
        \end{axis}
        \end{tikzpicture}
    \end{subfigure}
    \hfill
    \vspace{-3mm}
\begin{subfigure}[c]{\textwidth}
        \begin{tikzpicture}
        \begin{axis}[
                height=.18\textheight,
                width=\textwidth,
                ybar=0pt,
                ymax=7,
                ymin=0,
                clip=false,
                axis lines*=left,
                bar width=3.9pt,
                enlarge x limits=0.05,
                legend style={at={(0.55,1.3)}, anchor=east,legend columns=2, font=\footnotesize},
                ylabel={GFlop/s},
                xlabel={Matrices},
                symbolic x coords={Si87H76 (f64),Si87H76 (f32),spal (f64),spal (f32),torso1 (f64),torso1 (f32),TSOPF (f64),TSOPF (f32),wikipedia-20060925 (f64),wikipedia-20060925 (f32),average (f64), average (f32)},
                xticklabels={Si87H76 (f64),Si87H76 (f32),spal (f64),spal (f32),torso1 (f64),torso1 (f32),TSOPF (f64),TSOPF (f32),wikipedia-20060925 (f64),wikipedia-20060925 (f32),average (f64), average (f32)},
                xtick=data,
                nodes near coords,
                point meta=explicit symbolic,
                every node near coord/.append style={font=\footnotesize},
                yticklabel style = {font=\footnotesize,xshift=0.5ex},
                xticklabel style = {font=\footnotesize,yshift=0.5ex},
                ylabel style = {font=\footnotesize},
                xlabel style = {font=\footnotesize},
                every axis legend/.append style={nodes={right}},
                every node near coord/.append style={font=\tiny, rotate=90, anchor=west},
                x tick label style={rotate=50,anchor=east},
                ymajorgrids = true,
            ]

\addplot[draw=color_0!65!black, fill=color_0!30!white] coordinates {  (Si87H76 (f64), 0.357159)  (Si87H76 (f32), 0.361754)  (spal (f64), 0.367456)  (spal (f32), 0.394363)  (torso1 (f64), 0.410787)  (torso1 (f32), 0.416636)  (TSOPF (f64), 0.391946)  (TSOPF (f32), 0.394165)  (wikipedia-20060925 (f64), 0.127259)  (wikipedia-20060925 (f32), 0.145002)  (average (f64), 0.17822350000000003)  (average (f32), 0.18129254347826085)  };
\addplot[draw=color_1!65!black, fill=color_1!30!white] coordinates {  (Si87H76 (f64), 0.488201)[1.4]  (Si87H76 (f32), 0.505167)[1.4]  (spal (f64), 1.32572)[3.6]  (spal (f32), 2.35582)[6.0]  (torso1 (f64), 1.41814)[3.5]  (torso1 (f32), 1.92828)[4.6]  (TSOPF (f64), 2.46082)[6.3]  (TSOPF (f32), 4.21158)[10.7]  (wikipedia-20060925 (f64), 0.087967)[0.7]  (wikipedia-20060925 (f32), 0.102885)[0.7] (average (f64), 0.5274640652173912)[3.0]  (average (f32), 0.745644652173913)[4.1]   };
\addplot[draw=color_2!65!black, fill=color_2!30!white] coordinates {  (Si87H76 (f64), 0.637607)[1.8]  (Si87H76 (f32), 0.67444)[1.9]  (spal (f64), 1.13898)[3.1]  (spal (f32), 1.88205)[4.8]  (torso1 (f64), 1.94759)[4.7]  (torso1 (f32), 2.81674)[6.8]  (TSOPF (f64), 3.02269)[7.7]  (TSOPF (f32), 5.37968)[13.6]  (wikipedia-20060925 (f64), 0.0647835)[0.5]  (wikipedia-20060925 (f32), 0.0794812)[0.5]  (average (f64), 0.6444112717391305)[3.6]  (average (f32), 0.9354526999999998)[5.2]  };
\addplot[draw=color_3!65!black, fill=color_3!30!white] coordinates {  (Si87H76 (f64), 0.68992)[1.9]  (Si87H76 (f32), 0.747584)[2.1]  (spal (f64), 0.776781)[2.1]  (spal (f32), 1.40329)[3.6]  (torso1 (f64), 2.2317)[5.4]  (torso1 (f32), 3.19422)[7.7]  (TSOPF (f64), 3.14786)[8.0]  (TSOPF (f32), 5.73568)[14.6]  (wikipedia-20060925 (f64), 0.0474062)[0.4]  (wikipedia-20060925 (f32), 0.0572043)[0.4]  (average (f64), 0.6608736130434784)[3.7]  (average (f32), 1.0085277456521742)[5.6] };
\addplot[draw=color_4!65!black, fill=color_4!30!white] coordinates {  (Si87H76 (f64), 0.493337)[1.4]  (Si87H76 (f32), 0.565454)[1.6]  (spal (f64), 0.385535)[1.0]  (spal (f32), 0.69571)[1.8]  (torso1 (f64), 1.48046)[3.6]  (torso1 (f32), 2.65508)[6.4]  (TSOPF (f64), 2.33133)[5.9]  (TSOPF (f32), 4.47289)[11.3]  (wikipedia-20060925 (f64), 0.0261127)[0.2]  (wikipedia-20060925 (f32), 0.0301134)[0.2]  (average (f64), 0.46044753913043485)[2.6]  (average (f32), 0.7716340326086955)[4.3] };

        \end{axis}
        \end{tikzpicture}
    \end{subfigure}
        \caption{Performance in Giga Flop per second for sequential computation in double and single precision for our SPC5 kernels on Fujitsu-SVE architecture.
                 Speedup of SPC5 is computed against the scalar sequential version and written above the bars.
                 }
        \label{res:perf:full:sve}
\end{figure}
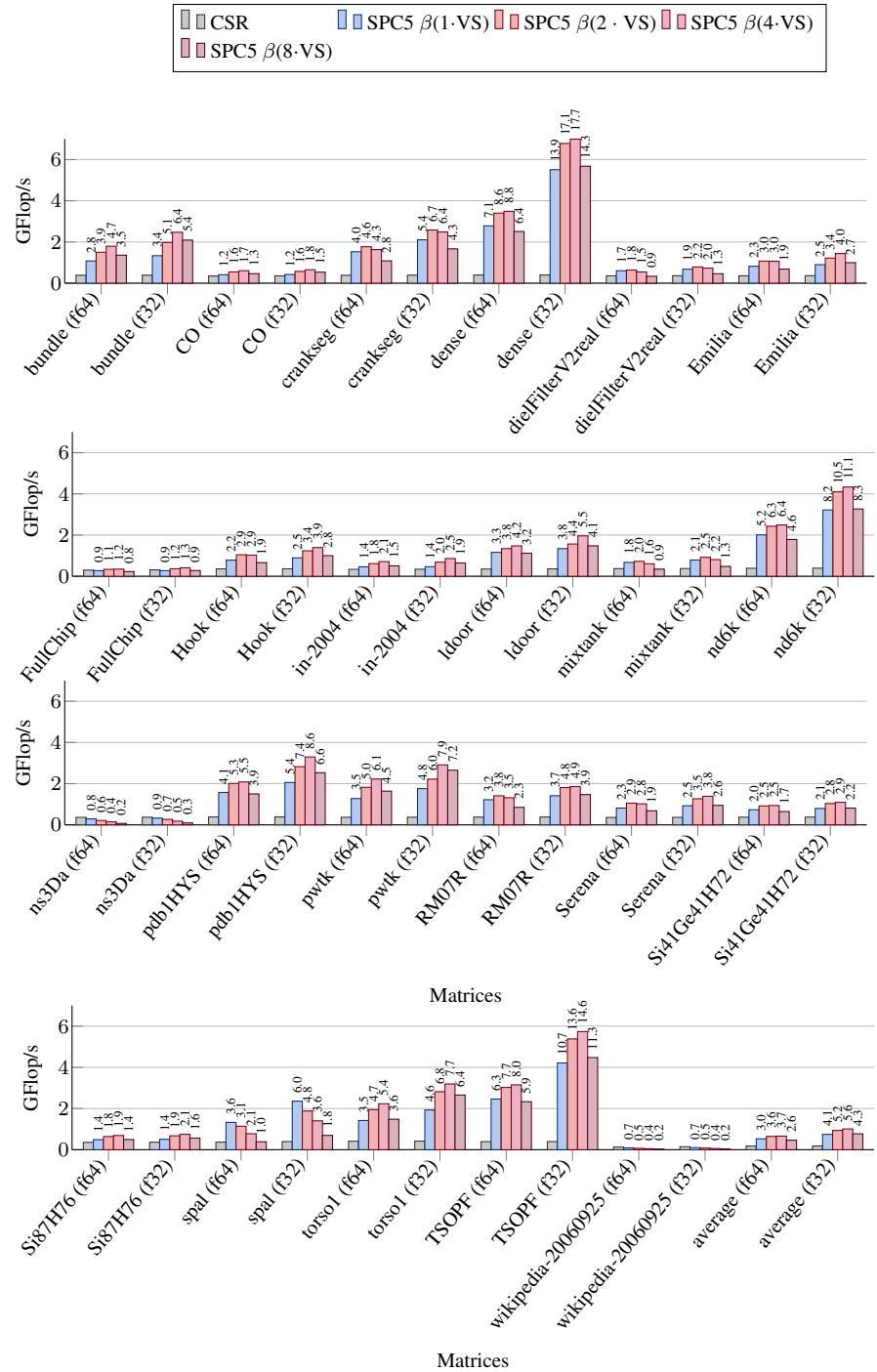

%% file: figcomparmavxtab.tex
\begin{table}[htbp]
\footnotesize
  \centering  
\begin{minipage}{\linewidth}
  \centering
\begin{tabular}{|c|c|c|c|c|c|c|c|c|c|c|c|} \hline
 & $x$ load / & \multicolumn{2}{c|}{CSR}& \multicolumn{2}{c|}{$\beta$(1,VS)}& \multicolumn{2}{c|}{$\beta$(2,VS)}& \multicolumn{2}{c|}{$\beta$(4,VS)}& \multicolumn{2}{c|}{$\beta$(8,VS)} \\ \cline{3-12}
 & reduction & f64 & f32 & f64 & f32 & f64 & f32 & f64 & f32 & f64 & f32  \\ \hline
\multirow{4}{*}{CO} & Yes/Yes & 0.3 & 0.4  & 0.4 [x1.2]  & 0.4 [x1.2]  & 0.5 [x1.6]  & 0.6 [x1.6]  & 0.6 [x1.7]  & 0.6 [x1.8]  & 0.5 [x1.3]  & 0.5 [x1.5] \\
 & Yes/No &  &  &  &  &  &  &  &  &  & 0.5 [x1.4] \\
 & No/Yes &  &  &  &  & 0.5 [x1.5]  &  & 0.5 [x1.5]  & 0.6 [x1.6]  & 0.5 [x1.5]  & 0.6 [x1.6] \\
 & No/No &  &  &  &  & 0.5 [x1.5]  &  & 0.5 [x1.5]  & 0.6 [x1.6]  & 0.5 [x1.5]  & 0.6 [x1.6] \\
\hline
\multirow{4}{*}{dense} & Yes/Yes & 0.4 & 0.4  & 2.8 [x7.1]  & 5.5 [x13.9]  & 3.4 [x8.6]  & 6.8 [x17.1]  & 3.5 [x8.8]  & 7.0 [x17.7]  & 2.5 [x6.4]  & 5.7 [x14.3] \\
 & Yes/No &  &  &  &  &  & 6.9 [x17.5]  &  & 6.5 [x16.4]  & 2.5 [x6.3]  & 6.4 [x16.0] \\
 & No/Yes &  &  & 2.8 [x7.0]  &  & 3.3 [x8.5]  & 6.8 [x17.2]  & 3.0 [x7.7]  & 6.4 [x16.1]  & 3.1 [x7.9]  & 6.4 [x16.1] \\
 & No/No &  &  & 2.8 [x7.0]  & 5.5 [x13.8]  & 3.3 [x8.5]  & 6.8 [x17.2]  & 3.0 [x7.7]  & 6.4 [x16.2]  & 3.1 [x7.9]  & 6.4 [x16.1] \\
\hline
\multirow{4}{*}{nd6k} & Yes/Yes & 0.4 & 0.4  & 2.0 [x5.2]  & 3.2 [x8.2]  & 2.4 [x6.3]  & 4.1 [x10.5]  & 2.5 [x6.4]  & 4.3 [x11.1]  & 1.8 [x4.6]  & 3.3 [x8.3] \\
 & Yes/No &  &  &  &  &  &  &  &  &  & \\
 & No/Yes &  &  &  &  &  &  & 2.2 [x5.8]  & 3.9 [x9.9]  & 2.0 [x5.2]  & 3.6 [x9.3] \\
 & No/No &  &  &  &  &  &  & 2.2 [x5.8]  & 3.9 [x9.9]  & 2.0 [x5.2]  & 3.6 [x9.3] \\
\hline
\multirow{4}{*}{average} & Yes/Yes & 0.2 & 0.2  & 0.5 [x3.0]  & 0.7 [x4.1]  & 0.6 [x3.6]  & 0.9 [x5.2]  & 0.7 [x3.7]  & 1.0 [x5.6]  & 0.5 [x2.6]  & 0.8 [x4.3] \\
 & Yes/No &  &  &  &  &  &  &  & 1.0 [x5.5]  &  & \\
 & No/Yes &  &  &  &  & 0.6 [x3.5]  & 0.9 [x5.0]  & 0.6 [x3.3]  & 0.9 [x4.8]  & 0.5 [x2.9]  & 0.8 [x4.6] \\
 & No/No &  &  &  &  & 0.6 [x3.5]  & 0.9 [x5.0]  & 0.6 [x3.3]  & 0.9 [x4.8]  & 0.5 [x2.9]  & 0.8 [x4.6] \\
\hline

  \end{tabular}
  \caption*{(a) Fujitsu-SVE}
\end{minipage}

\begin{minipage}{\linewidth}
  \centering
\begin{tabular}{|c|c|c|c|c|c|c|c|c|c|c|c|c|c|} \hline
 & $x$ load / & \multicolumn{2}{c|}{CSR}& \multicolumn{2}{c|}{MKL}& \multicolumn{2}{c|}{$\beta$(1,VS)}& \multicolumn{2}{c|}{$\beta$(2,VS)}& \multicolumn{2}{c|}{$\beta$(4,VS)}& \multicolumn{2}{c|}{$\beta$(8,VS)} \\ \cline{3-14}
 & reduction & f64 & f32 & f64 & f32 & f64 & f32 & f64 & f32 & f64 & f32 & f64 & f32  \\ \hline
\multirow{2}{*}{CO} & No/Yes & 1.4 & 1.9  & 1.9 [x1.3]  & 2.3 [x1.2]  & 1.3 [x0.9]  & 1.4 [x0.8]  & 1.6 [x1.1]  & 1.9 [x1.0]  & 1.7 [x1.2]  & 1.9 [x1.0]  & 1.6 [x1.1]  & 1.9 [x1.0] \\
 & No/No &  &  &  &  &  &  &  &  &  &  &  & \\
\hline
\multirow{2}{*}{dense} & No/Yes & 1.2 & 1.3  & 2.3 [x1.9]  & 3.6 [x2.8]  & 3.7 [x3.0]  & 8.3 [x6.4]  & 4.1 [x3.4]  & 9.5 [x7.3]  & 4.3 [x3.6]  & 11.2 [x8.6]  & 4.4 [x3.6]  & 10.8 [x8.3] \\
 & No/No &  &  &  &  &  & 9.4 [x7.2]  &  & 10.6 [x8.1]  & 4.2 [x3.4]  & 11.0 [x8.5]  &  & 11.0 [x8.5] \\
\hline
\multirow{2}{*}{nd6k} & No/Yes & 1.2 & 1.4  & 2.2 [x1.8]  & 2.8 [x2.0]  & 2.9 [x2.4]  & 6.2 [x4.5]  & 3.4 [x2.8]  & 7.3 [x5.4]  & 3.4 [x2.8]  & 7.4 [x5.4]  & 3.4 [x2.8]  & 7.4 [x5.4] \\
 & No/No &  &  &  & 2.8 [x2.1]  &  & 6.3 [x4.6]  &  & 7.3 [x5.3]  &  & 7.6 [x5.6]  &  & 7.1 [x5.2] \\
\hline
\multirow{2}{*}{average} & No/Yes & 0.6 & 0.8  & 0.9 [x1.5]  & 1.2 [x1.6]  & 1.0 [x1.6]  & 1.7 [x2.3]  & 1.2 [x1.8]  & 2.0 [x2.6]  & 1.2 [x1.8]  & 2.0 [x2.7]  & 1.1 [x1.7]  & 1.9 [x2.5] \\
 & No/No &  &  &  &  & 1.1 [x1.7]  & 1.8 [x2.4]  &  &  &  & 2.0 [x2.6]  &  & 1.8 [x2.4] \\
\hline

  \end{tabular}
  \caption*{(b) Intel-AVX512}
\end{minipage}
  
  \caption{Performance in Giga Flop per second for sequential computation in double and single precision for our SPC5 kernels on Fujitsu-SVE and Intel-AVX512 architectures.
                 Speedup of SPC5 is computed against the scalar sequential version, we print the values only when there is a difference with the first version (above one digit difference in the speedup).
                 We provide the results for the CO, dense and nd6k matrices, and the average based on all the matrices from the test set.
                 We compare the the loading of full $x$ vectors per block (SVE and AVX512), and the use of manual multi-reduction against vectorial reduction (SVE only).
                 The scalar and $\beta$(1,VS) and MKL versions are expected to remain unchanged, differences for these kernels are from noise.}
  \label{res:perf:comp}
\end{table}

%% file: figplotavx.tex
\begin{figure}[h!]
\centering
    \begin{subfigure}[c]{\textwidth}
    \begin{tikzpicture}
    \begin{axis}[
            height=.25\textheight,
            width=\textwidth,
            ymin=0,
            xmin=0,
            clip=false,
            axis lines*=left,
            legend style={at={(0.9,1.3)}, anchor=east,legend columns=4, font=\footnotesize},
            ylabel={GFlop/s},
            xlabel={Average NNZ per blocks},
            yticklabel style = {font=\footnotesize,xshift=0.5ex},
            xticklabel style = {font=\footnotesize,yshift=0.5ex},
            ylabel style = {font=\footnotesize},
            xlabel style = {font=\footnotesize},
    every node near coord/.style={font=\tiny}
        ]
\addplot[draw=color_1, fill=color_1!30!white, only marks] table[x index=8, y index=9] {dataavx.txt};
\addplot[draw=color_2, fill=color_2!30!white, only marks] table[x index=10, y index=11] {dataavx.txt};
\addplot[draw=color_3, fill=color_3!30!white, only marks] table[x index=12, y index=13] {dataavx.txt};
\addplot[draw=color_4, fill=color_4!30!white, only marks] table[x index=14, y index=15] {dataavx.txt};

 \legend{SPC5 $\beta$(1$\cdot$VS), SPC5 $\beta$(2$\cdot$VS), SPC5 $\beta$(4$\cdot$VS), SPC5 $\beta$(8$\cdot$VS)}
        \end{axis}
        \end{tikzpicture}
        \caption{Double precision (f64).}
        \label{res:plotavx:f64}
    \end{subfigure}
    \hfill
    \vspace{-3mm}
    \begin{subfigure}[c]{\textwidth}
    \begin{tikzpicture}
    \begin{axis}[
            height=.25\textheight,
            width=\textwidth,
            ymin=0,
            xmin=0,
            clip=false,
            axis lines*=left,
            legend style={at={(0.9,1.3)}, anchor=east,legend columns=4, font=\footnotesize},
            ylabel={GFlop/s},
            xlabel={Average NNZ per blocks},
            yticklabel style = {font=\footnotesize,xshift=0.5ex},
            xticklabel style = {font=\footnotesize,yshift=0.5ex},
            ylabel style = {font=\footnotesize},
            xlabel style = {font=\footnotesize},
    every node near coord/.style={font=\tiny}
        ]
\addplot[draw=color_1, fill=color_1!30!white, only marks] table[x index=0, y index=1] {dataavx.txt};
\addplot[draw=color_2, fill=color_2!30!white, only marks] table[x index=2, y index=3] {dataavx.txt};
\addplot[draw=color_3, fill=color_3!30!white, only marks] table[x index=4, y index=5] {dataavx.txt};
\addplot[draw=color_4, fill=color_4!30!white, only marks] table[x index=6, y index=7] {dataavx.txt};

        \end{axis}
        \end{tikzpicture}
        \caption{Single precision (f32).}
        \label{res:plotavx:f32}
    \end{subfigure}
        \caption{Performance in Giga Flop per second for sequential computation in double and single precision for our SPC5 kernels on Intel-AVX architecture for all the matrices of the test set.}
        \label{res:plotavx}
\end{figure}
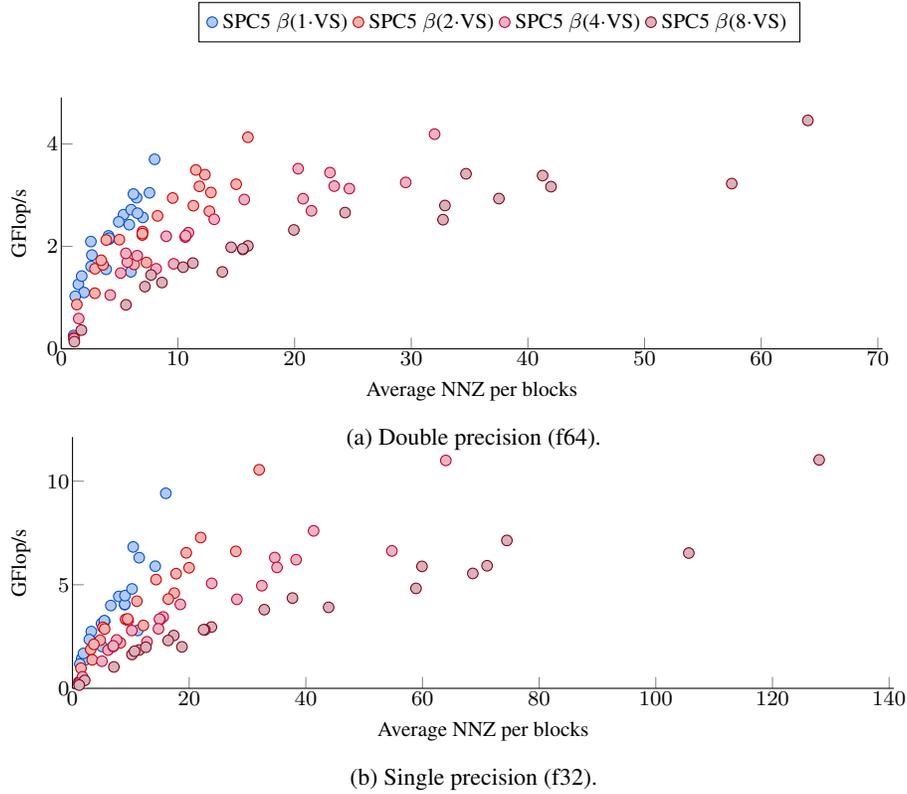

%% file: figresavx.tex
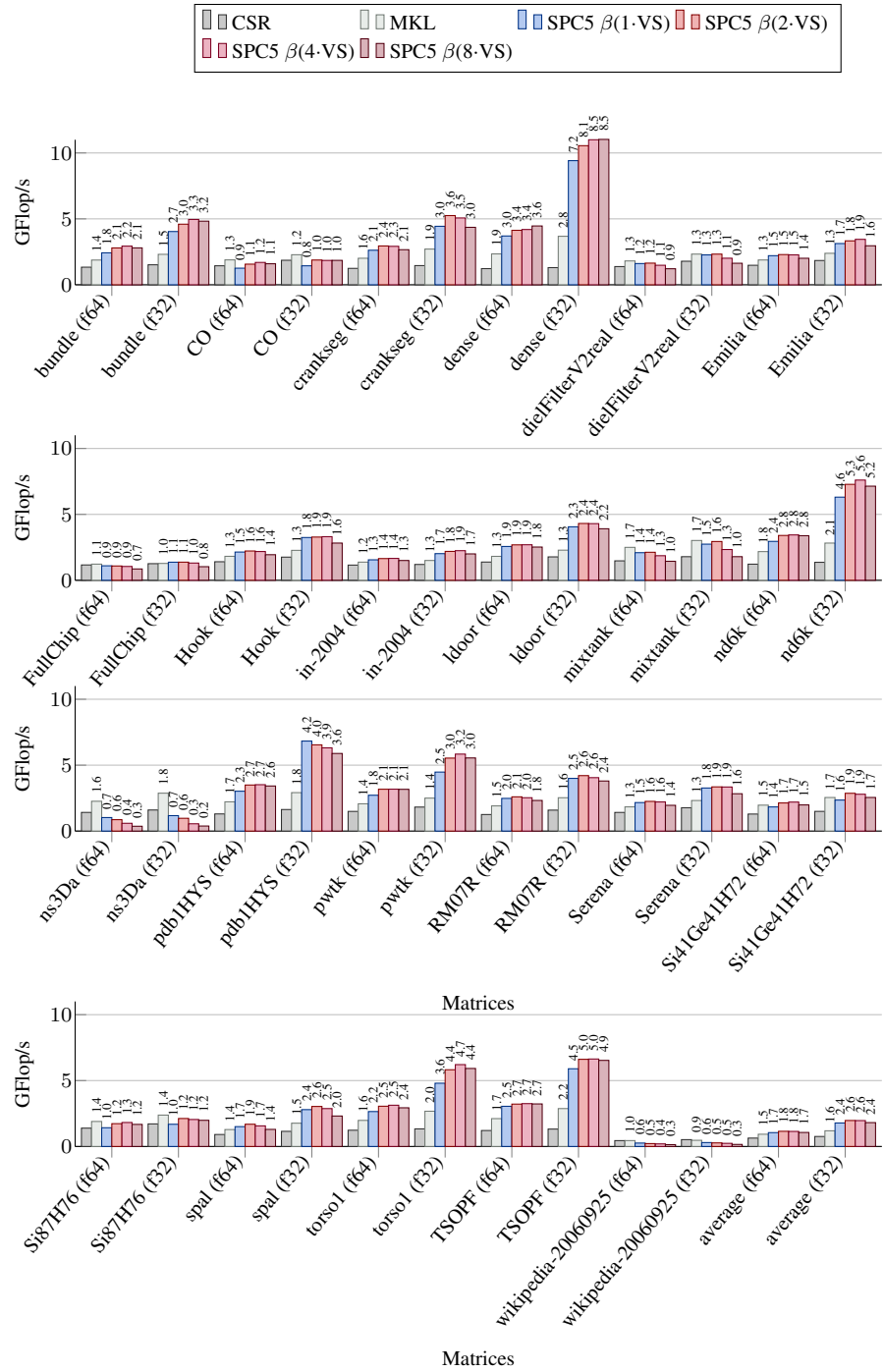
\begin{figure}[h!]
\centering
    \begin{subfigure}[c]{\textwidth}
        \begin{tikzpicture}
        \begin{axis}[
                height=.18\textheight,
                width=\textwidth,
                ybar=0pt,
                ymax=11,
                ymin=0,
                clip=false,
                axis lines*=left,
                bar width=3.9pt,
                enlarge x limits=0.05,
                legend style={at={(0.95,1.7)}, anchor=east,legend columns=4, font=\footnotesize},
                ylabel={GFlop/s},
                symbolic x coords={bundle (f64),bundle (f32),CO (f64),CO (f32),crankseg (f64),crankseg (f32),dense (f64),dense (f32),dielFilterV2real (f64),dielFilterV2real (f32),Emilia (f64),Emilia (f32)},
                xticklabels={bundle (f64),bundle (f32),CO (f64),CO (f32),crankseg (f64),crankseg (f32),dense (f64),dense (f32),dielFilterV2real (f64),dielFilterV2real (f32),Emilia (f64),Emilia (f32)},
                xtick=data,
                nodes near coords,
                point meta=explicit symbolic,
                every node near coord/.append style={font=\footnotesize},
                nodes near coords align={vertical},
                yticklabel style = {font=\footnotesize,xshift=0.5ex},
                xticklabel style = {font=\footnotesize,yshift=0.5ex},
                ylabel style = {font=\footnotesize},
                xlabel style = {font=\footnotesize},
                every axis legend/.append style={nodes={right}},
                every node near coord/.append style={font=\tiny, rotate=90, anchor=west},
                x tick label style={rotate=50,anchor=east},
                ymajorgrids = true,
            ]
\addplot[draw=color_0!65!black, fill=color_0!30!white] coordinates {  (bundle (f64), 1.33522)  (bundle (f32), 1.51641)  (CO (f64), 1.44112)  (CO (f32), 1.85988)  (crankseg (f64), 1.24501)  (crankseg (f32), 1.4559)  (dense (f64), 1.22599)  (dense (f32), 1.30155)  (dielFilterV2real (f64), 1.38076)  (dielFilterV2real (f32), 1.78603)  (Emilia (f64), 1.47917)  (Emilia (f32), 1.84082)  };
\addplot[draw=color_mkl!65!black, fill=color_mkl!30!white] coordinates {  (bundle (f64), 1.87605)[1.4]  (bundle (f32), 2.30932)[1.5]  (CO (f64), 1.88775)[1.3]  (CO (f32), 2.2788)[1.2]  (crankseg (f64), 2.01268)[1.6]  (crankseg (f32), 2.70841)[1.9]  (dense (f64), 2.34221)[1.9]  (dense (f32), 3.67929)[2.8]  (dielFilterV2real (f64), 1.80851)[1.3]  (dielFilterV2real (f32), 2.3267)[1.3]  (Emilia (f64), 1.8811)[1.3]  (Emilia (f32), 2.38783)[1.3]  };
\addplot[draw=color_1!65!black, fill=color_1!30!white] coordinates {  (bundle (f64), 2.42434)[1.8]  (bundle (f32), 4.03994)[2.7]  (CO (f64), 1.2597)[0.9]  (CO (f32), 1.43904)[0.8]  (crankseg (f64), 2.62064)[2.1]  (crankseg (f32), 4.43682)[3.0]  (dense (f64), 3.7017)[3.0]  (dense (f32), 9.41655)[7.2]  (dielFilterV2real (f64), 1.60964)[1.2]  (dielFilterV2real (f32), 2.26678)[1.3]  (Emilia (f64), 2.20698)[1.5]  (Emilia (f32), 3.1282)[1.7]  };
\addplot[draw=color_2!65!black, fill=color_2!30!white] coordinates {  (bundle (f64), 2.79602)[2.1]  (bundle (f32), 4.59641)[3.0]  (CO (f64), 1.56409)[1.1]  (CO (f32), 1.87748)[1.0]  (crankseg (f64), 2.94589)[2.4]  (crankseg (f32), 5.25063)[3.6]  (dense (f64), 4.13209)[3.4]  (dense (f32), 10.5522)[8.1]  (dielFilterV2real (f64), 1.64121)[1.2]  (dielFilterV2real (f32), 2.3251)[1.3]  (Emilia (f64), 2.29087)[1.5]  (Emilia (f32), 3.33204)[1.8]  };
\addplot[draw=color_3!65!black, fill=color_3!30!white] coordinates {  (bundle (f64), 2.93175)[2.2]  (bundle (f32), 4.95563)[3.3]  (CO (f64), 1.69542)[1.2]  (CO (f32), 1.84742)[1.0]  (crankseg (f64), 2.91633)[2.3]  (crankseg (f32), 5.06876)[3.5]  (dense (f64), 4.19328)[3.4]  (dense (f32), 11.0021)[8.5]  (dielFilterV2real (f64), 1.48035)[1.1]  (dielFilterV2real (f32), 2.01501)[1.1]  (Emilia (f64), 2.2685)[1.5]  (Emilia (f32), 3.44648)[1.9]  };
\addplot[draw=color_4!65!black, fill=color_4!30!white] coordinates {  (bundle (f64), 2.79932)[2.1]  (bundle (f32), 4.82506)[3.2]  (CO (f64), 1.59471)[1.1]  (CO (f32), 1.85132)[1.0]  (crankseg (f64), 2.66104)[2.1]  (crankseg (f32), 4.35851)[3.0]  (dense (f64), 4.46116)[3.6]  (dense (f32), 11.029)[8.5]  (dielFilterV2real (f64), 1.21583)[0.9]  (dielFilterV2real (f32), 1.63243)[0.9]  (Emilia (f64), 2.01085)[1.4]  (Emilia (f32), 2.95864)[1.6]  };

 \legend{CSR, MKL, SPC5 $\beta$(1$\cdot$VS), SPC5 $\beta$(2$\cdot$VS), SPC5 $\beta$(4$\cdot$VS), SPC5 $\beta$(8$\cdot$VS)}
        \end{axis}
        \end{tikzpicture}
    \end{subfigure}
    \hfill
    \vspace{-3mm}
    \begin{subfigure}[c]{\textwidth}
        \begin{tikzpicture}
        \begin{axis}[
                height=.18\textheight,
                width=\textwidth,
                ybar=0pt,
                ymax=11,
                ymin=0,
                clip=false,
                axis lines*=left,
                bar width=3.9pt,
                enlarge x limits=0.05,
                legend style={at={(0.95,1.3)}, anchor=east,legend columns=6, font=\footnotesize},
                ylabel={GFlop/s},
                symbolic x coords={FullChip (f64),FullChip (f32),Hook (f64),Hook (f32),in-2004 (f64),in-2004 (f32),ldoor (f64),ldoor (f32),mixtank (f64),mixtank (f32),nd6k (f64),nd6k (f32)},
                xticklabels={FullChip (f64),FullChip (f32),Hook (f64),Hook (f32),in-2004 (f64),in-2004 (f32),ldoor (f64),ldoor (f32),mixtank (f64),mixtank (f32),nd6k (f64),nd6k (f32)},
                xtick=data,
                nodes near coords,
                point meta=explicit symbolic,
                every node near coord/.append style={font=\footnotesize},
                yticklabel style = {font=\footnotesize,xshift=0.5ex},
                xticklabel style = {font=\footnotesize,yshift=0.5ex},
                ylabel style = {font=\footnotesize},
                xlabel style = {font=\footnotesize},
                every axis legend/.append style={nodes={right}},
                every node near coord/.append style={font=\tiny, rotate=90, anchor=west},
                x tick label style={rotate=50,anchor=east},
                ymajorgrids = true,
            ]

\addplot[draw=color_0!65!black, fill=color_0!30!white] coordinates {  (FullChip (f64), 1.16359)  (FullChip (f32), 1.26166)  (Hook (f64), 1.40813)  (Hook (f32), 1.75748)  (in-2004 (f64), 1.15331)  (in-2004 (f32), 1.20169)  (ldoor (f64), 1.3844)  (ldoor (f32), 1.77669)  (mixtank (f64), 1.47594)  (mixtank (f32), 1.79112)  (nd6k (f64), 1.22359)  (nd6k (f32), 1.36538)  };
\addplot[draw=color_mkl!65!black, fill=color_mkl!30!white] coordinates {  (FullChip (f64), 1.22635)[1.1]  (FullChip (f32), 1.27998)[1.0]  (Hook (f64), 1.81922)[1.3]  (Hook (f32), 2.27189)[1.3]  (in-2004 (f64), 1.37807)[1.2]  (in-2004 (f32), 1.50437)[1.3]  (ldoor (f64), 1.81667)[1.3]  (ldoor (f32), 2.28538)[1.3]  (mixtank (f64), 2.49874)[1.7]  (mixtank (f32), 3.02054)[1.7]  (nd6k (f64), 2.17754)[1.8]  (nd6k (f32), 2.82372)[2.1]  };
\addplot[draw=color_1!65!black, fill=color_1!30!white] coordinates {  (FullChip (f64), 1.10047)[0.9]  (FullChip (f32), 1.3731)[1.1]  (Hook (f64), 2.14319)[1.5]  (Hook (f32), 3.24557)[1.8]  (in-2004 (f64), 1.55357)[1.3]  (in-2004 (f32), 2.02221)[1.7]  (ldoor (f64), 2.5664)[1.9]  (ldoor (f32), 4.05806)[2.3]  (mixtank (f64), 2.0955)[1.4]  (mixtank (f32), 2.74198)[1.5]  (nd6k (f64), 2.95247)[2.4]  (nd6k (f32), 6.30739)[4.6]  };
\addplot[draw=color_2!65!black, fill=color_2!30!white] coordinates {  (FullChip (f64), 1.08654)[0.9]  (FullChip (f32), 1.3763)[1.1]  (Hook (f64), 2.22342)[1.6]  (Hook (f32), 3.2799)[1.9]  (in-2004 (f64), 1.64853)[1.4]  (in-2004 (f32), 2.19294)[1.8]  (ldoor (f64), 2.68833)[1.9]  (ldoor (f32), 4.31095)[2.4]  (mixtank (f64), 2.12582)[1.4]  (mixtank (f32), 2.9429)[1.6]  (nd6k (f64), 3.40169)[2.8]  (nd6k (f32), 7.27939)[5.3]  };
\addplot[draw=color_3!65!black, fill=color_3!30!white] coordinates {  (FullChip (f64), 1.05068)[0.9]  (FullChip (f32), 1.30871)[1.0]  (Hook (f64), 2.18302)[1.6]  (Hook (f32), 3.30978)[1.9]  (in-2004 (f64), 1.66102)[1.4]  (in-2004 (f32), 2.24796)[1.9]  (ldoor (f64), 2.6957)[1.9]  (ldoor (f32), 4.29826)[2.4]  (mixtank (f64), 1.86536)[1.3]  (mixtank (f32), 2.33561)[1.3]  (nd6k (f64), 3.44216)[2.8]  (nd6k (f32), 7.6028)[5.6]  };
\addplot[draw=color_4!65!black, fill=color_4!30!white] coordinates {  (FullChip (f64), 0.859694)[0.7]  (FullChip (f32), 1.03303)[0.8]  (Hook (f64), 1.94112)[1.4]  (Hook (f32), 2.82388)[1.6]  (in-2004 (f64), 1.50231)[1.3]  (in-2004 (f32), 2.00274)[1.7]  (ldoor (f64), 2.52252)[1.8]  (ldoor (f32), 3.91011)[2.2]  (mixtank (f64), 1.44285)[1.0]  (mixtank (f32), 1.79274)[1.0]  (nd6k (f64), 3.38394)[2.8]  (nd6k (f32), 7.14043)[5.2]  };

        \end{axis}
        \end{tikzpicture}
    \end{subfigure}
    \hfill
    \vspace{-3mm}
\begin{subfigure}[c]{\textwidth}
        \begin{tikzpicture}
        \begin{axis}[
                height=.18\textheight,
                width=\textwidth,
                ybar=0pt,
                ymax=11,
                ymin=0,
                clip=false,
                axis lines*=left,
                bar width=3.9pt,
                enlarge x limits=0.05,
                legend style={at={(0.55,1.3)}, anchor=east,legend columns=2, font=\footnotesize},
                ylabel={GFlop/s},
                xlabel={Matrices},
                symbolic x coords={ns3Da (f64),ns3Da (f32),pdb1HYS (f64),pdb1HYS (f32),pwtk (f64),pwtk (f32),RM07R (f64),RM07R (f32),Serena (f64),Serena (f32),Si41Ge41H72 (f64),Si41Ge41H72 (f32)},
                xticklabels={ns3Da (f64),ns3Da (f32),pdb1HYS (f64),pdb1HYS (f32),pwtk (f64),pwtk (f32),RM07R (f64),RM07R (f32),Serena (f64),Serena (f32),Si41Ge41H72 (f64),Si41Ge41H72 (f32)},
                xtick=data,
                nodes near coords,
                point meta=explicit symbolic,
                every node near coord/.append style={font=\footnotesize},
                yticklabel style = {font=\footnotesize,xshift=0.5ex},
                xticklabel style = {font=\footnotesize,yshift=0.5ex},
                ylabel style = {font=\footnotesize},
                xlabel style = {font=\footnotesize},
                every axis legend/.append style={nodes={right}},
                every node near coord/.append style={font=\tiny, rotate=90, anchor=west},
                x tick label style={rotate=50,anchor=east},
                ymajorgrids = true,
            ]

\addplot[draw=color_0!65!black, fill=color_0!30!white] coordinates {  (ns3Da (f64), 1.41532)  (ns3Da (f32), 1.59911)  (pdb1HYS (f64), 1.30113)  (pdb1HYS (f32), 1.63448)  (pwtk (f64), 1.49192)  (pwtk (f32), 1.82091)  (RM07R (f64), 1.25555)  (RM07R (f32), 1.58846)  (Serena (f64), 1.41393)  (Serena (f32), 1.76918)  (Si41Ge41H72 (f64), 1.29142)  (Si41Ge41H72 (f32), 1.49466)  };
\addplot[draw=color_mkl!65!black, fill=color_mkl!30!white] coordinates {  (ns3Da (f64), 2.26163)[1.6]  (ns3Da (f32), 2.87415)[1.8]  (pdb1HYS (f64), 2.21901)[1.7]  (pdb1HYS (f32), 2.91648)[1.8]  (pwtk (f64), 2.06803)[1.4]  (pwtk (f32), 2.50057)[1.4]  (RM07R (f64), 1.90977)[1.5]  (RM07R (f32), 2.52912)[1.6]  (Serena (f64), 1.83674)[1.3]  (Serena (f32), 2.318)[1.3]  (Si41Ge41H72 (f64), 1.96634)[1.5]  (Si41Ge41H72 (f32), 2.54622)[1.7]  };
\addplot[draw=color_1!65!black, fill=color_1!30!white] coordinates {  (ns3Da (f64), 1.02602)[0.7]  (ns3Da (f32), 1.17232)[0.7]  (pdb1HYS (f64), 3.02358)[2.3]  (pdb1HYS (f32), 6.82764)[4.2]  (pwtk (f64), 2.71831)[1.8]  (pwtk (f32), 4.47517)[2.5]  (RM07R (f64), 2.47972)[2.0]  (RM07R (f32), 3.99795)[2.5]  (Serena (f64), 2.16215)[1.5]  (Serena (f32), 3.26698)[1.8]  (Si41Ge41H72 (f64), 1.83147)[1.4]  (Si41Ge41H72 (f32), 2.35235)[1.6]  };
\addplot[draw=color_2!65!black, fill=color_2!30!white] coordinates {  (ns3Da (f64), 0.864853)[0.6]  (ns3Da (f32), 0.970372)[0.6]  (pdb1HYS (f64), 3.49349)[2.7]  (pdb1HYS (f32), 6.5418)[4.0]  (pwtk (f64), 3.17401)[2.1]  (pwtk (f32), 5.53548)[3.0]  (RM07R (f64), 2.59693)[2.1]  (RM07R (f32), 4.20649)[2.6]  (Serena (f64), 2.25165)[1.6]  (Serena (f32), 3.35278)[1.9]  (Si41Ge41H72 (f64), 2.1326)[1.7]  (Si41Ge41H72 (f32), 2.86534)[1.9]  };
\addplot[draw=color_3!65!black, fill=color_3!30!white] coordinates {  (ns3Da (f64), 0.591441)[0.4]  (ns3Da (f32), 0.553995)[0.3]  (pdb1HYS (f64), 3.51835)[2.7]  (pdb1HYS (f32), 6.31419)[3.9]  (pwtk (f64), 3.17847)[2.1]  (pwtk (f32), 5.83821)[3.2]  (RM07R (f64), 2.52963)[2.0]  (RM07R (f32), 4.052)[2.6]  (Serena (f64), 2.20909)[1.6]  (Serena (f32), 3.34334)[1.9]  (Si41Ge41H72 (f64), 2.19889)[1.7]  (Si41Ge41H72 (f32), 2.79319)[1.9]  };
\addplot[draw=color_4!65!black, fill=color_4!30!white] coordinates {  (ns3Da (f64), 0.365911)[0.3]  (ns3Da (f32), 0.383201)[0.2]  (pdb1HYS (f64), 3.41941)[2.6]  (pdb1HYS (f32), 5.88676)[3.6]  (pwtk (f64), 3.16762)[2.1]  (pwtk (f32), 5.54713)[3.0]  (RM07R (f64), 2.32225)[1.8]  (RM07R (f32), 3.79709)[2.4]  (Serena (f64), 1.95338)[1.4]  (Serena (f32), 2.83212)[1.6]  (Si41Ge41H72 (f64), 1.98459)[1.5]  (Si41Ge41H72 (f32), 2.54941)[1.7]  };

        \end{axis}
        \end{tikzpicture}
    \end{subfigure}
    \hfill
    \vspace{-3mm}
\begin{subfigure}[c]{\textwidth}
        \begin{tikzpicture}
        \begin{axis}[
                height=.18\textheight,
                width=\textwidth,
                ybar=0pt,
                ymax=11,
                ymin=0,
                clip=false,
                axis lines*=left,
                bar width=3.9pt,
                enlarge x limits=0.05,
                legend style={at={(0.55,1.3)}, anchor=east,legend columns=2, font=\footnotesize},
                ylabel={GFlop/s},
                xlabel={Matrices},
                symbolic x coords={Si87H76 (f64),Si87H76 (f32),spal (f64),spal (f32),torso1 (f64),torso1 (f32),TSOPF (f64),TSOPF (f32),wikipedia-20060925 (f64),wikipedia-20060925 (f32),average (f64), average (f32)},
                xticklabels={Si87H76 (f64),Si87H76 (f32),spal (f64),spal (f32),torso1 (f64),torso1 (f32),TSOPF (f64),TSOPF (f32),wikipedia-20060925 (f64),wikipedia-20060925 (f32),average (f64), average (f32)},
                xtick=data,
                nodes near coords,
                point meta=explicit symbolic,
                every node near coord/.append style={font=\footnotesize},
                yticklabel style = {font=\footnotesize,xshift=0.5ex},
                xticklabel style = {font=\footnotesize,yshift=0.5ex},
                ylabel style = {font=\footnotesize},
                xlabel style = {font=\footnotesize},
                every axis legend/.append style={nodes={right}},
                every node near coord/.append style={font=\tiny, rotate=90, anchor=west},
                x tick label style={rotate=50,anchor=east},
                ymajorgrids = true,
            ]

\addplot[draw=color_0!65!black, fill=color_0!30!white] coordinates {  (Si87H76 (f64), 1.39166)  (Si87H76 (f32), 1.71205)  (spal (f64), 0.908298)  (spal (f32), 1.14823)  (torso1 (f64), 1.23188)  (torso1 (f32), 1.33377)  (TSOPF (f64), 1.21054)  (TSOPF (f32), 1.31978)  (wikipedia-20060925 (f64), 0.444534)  (wikipedia-20060925 (f32), 0.518598)  (average (f64), 0.6363567826086957)  (average (f32), 0.7576921304347826)  };
\addplot[draw=color_mkl!65!black, fill=color_mkl!30!white] coordinates {  (Si87H76 (f64), 1.89799)[1.4]  (Si87H76 (f32), 2.38008)[1.4]  (spal (f64), 1.28332)[1.4]  (spal (f32), 1.76663)[1.5]  (torso1 (f64), 1.9791)[1.6]  (torso1 (f32), 2.67311)[2.0]  (TSOPF (f64), 2.11346)[1.7]  (TSOPF (f32), 2.87542)[2.2]  (wikipedia-20060925 (f64), 0.445959)[1.0]  (wikipedia-20060925 (f32), 0.468519)[0.9]  (average (f64), 0.9283965000000002)[1.5]  (average (f32), 1.1896636739130435)[1.6]  };
\addplot[draw=color_1!65!black, fill=color_1!30!white] coordinates {  (Si87H76 (f64), 1.42145)[1.0]  (Si87H76 (f32), 1.68513)[1.0]  (spal (f64), 1.50476)[1.7]  (spal (f32), 2.79512)[2.4]  (torso1 (f64), 2.64929)[2.2]  (torso1 (f32), 4.79925)[3.6]  (TSOPF (f64), 3.04795)[2.5]  (TSOPF (f32), 5.89601)[4.5]  (wikipedia-20060925 (f64), 0.25957)[0.6]  (wikipedia-20060925 (f32), 0.297995)[0.6]  (average (f64), 1.0512797826086955)[1.7]  (average (f32), 1.7835120652173913)[2.4]  };
\addplot[draw=color_2!65!black, fill=color_2!30!white] coordinates {  (Si87H76 (f64), 1.73043)[1.2]  (Si87H76 (f32), 2.12215)[1.2]  (spal (f64), 1.68663)[1.9]  (spal (f32), 3.03602)[2.6]  (torso1 (f64), 3.05377)[2.5]  (torso1 (f32), 5.82099)[4.4]  (TSOPF (f64), 3.21435)[2.7]  (TSOPF (f32), 6.613)[5.0]  (wikipedia-20060925 (f64), 0.219669)[0.5]  (wikipedia-20060925 (f32), 0.271975)[0.5]  (average (f64), 1.1513670000000003)[1.8]  (average (f32), 1.9707094999999997)[2.6]  };
\addplot[draw=color_3!65!black, fill=color_3!30!white] coordinates {  (Si87H76 (f64), 1.81917)[1.3]  (Si87H76 (f32), 2.04718)[1.2]  (spal (f64), 1.56357)[1.7]  (spal (f32), 2.87468)[2.5]  (torso1 (f64), 3.12921)[2.5]  (torso1 (f32), 6.21062)[4.7]  (TSOPF (f64), 3.2504)[2.7]  (TSOPF (f32), 6.63533)[5.0]  (wikipedia-20060925 (f64), 0.197038)[0.4]  (wikipedia-20060925 (f32), 0.238752)[0.5]  (average (f64), 1.1428006304347824)[1.8]  (average (f32), 1.9639131956521743)[2.6]  };
\addplot[draw=color_4!65!black, fill=color_4!30!white] coordinates {  (Si87H76 (f64), 1.67484)[1.2]  (Si87H76 (f32), 1.98501)[1.2]  (spal (f64), 1.2957)[1.4]  (spal (f32), 2.31021)[2.0]  (torso1 (f64), 2.93466)[2.4]  (torso1 (f32), 5.91851)[4.4]  (TSOPF (f64), 3.22744)[2.7]  (TSOPF (f32), 6.53163)[4.9]  (wikipedia-20060925 (f64), 0.140278)[0.3]  (wikipedia-20060925 (f32), 0.156124)[0.3]  (average (f64), 1.062639630434783)[1.7]  (average (f32), 1.8098931521739132)[2.4]  };

        \end{axis}
        \end{tikzpicture}
    \end{subfigure}
        \caption{Performance in Giga Flop per second for sequential computation in double and single precision for our SPC5 kernels on Intel-AVX512 architecture.
                 Speedup of SPC5 is computed against the scalar sequential version and written above the bars.
                 }
        \label{res:perf:full:avx}
\end{figure}

%% file: figrespar.tex
\begin{figure}[h!]
\centering
    \begin{subfigure}[c]{\textwidth}
        \begin{tikzpicture}
        \begin{axis}[
                height=.18\textheight,
                width=\textwidth,
                ybar=0pt,
                ymin=0,
                clip=false,
                axis lines*=left,
                bar width=3.9pt,
                enlarge x limits=0.05,
                legend style={at={(0.95,1.7)}, anchor=east,legend columns=4, font=\footnotesize},
                ylabel={GFlop/s},
                symbolic x coords={CO (f64),CO (f32),dense (f64),dense (f32),nd6k (f64),nd6k (f32),average (f64),average (f32)},
                xticklabels={CO (f64),CO (f32),dense (f64),dense (f32),nd6k (f64),nd6k (f32),average (f64),average (f32)},
                xtick=data,
                nodes near coords,
                point meta=explicit symbolic,
                every node near coord/.append style={font=\footnotesize},
                nodes near coords align={vertical},
                yticklabel style = {font=\footnotesize,xshift=0.5ex},
                xticklabel style = {font=\footnotesize,yshift=0.5ex},
                ylabel style = {font=\footnotesize},
                xlabel style = {font=\footnotesize},
                every axis legend/.append style={nodes={right}},
                every node near coord/.append style={font=\tiny, rotate=90, anchor=west},
                x tick label style={rotate=50,anchor=east},
                ymajorgrids = true,
            ]
\addplot[draw=color_1!65!black, fill=color_1!30!white] coordinates {  (CO (f64), 17.5883)[42.9]  (CO (f32), 17.4754)[41.3]  (dense (f64), 89.2106)[32.1]  (dense (f32), 147.264)[26.9]  (nd6k (f64), 78.3996)[39.1]  (nd6k (f32), 119.529)[37.2]  (average (f64), 19.217891956521736)[36.6]  (average (f32), 25.867125)[34.7]  };
\addplot[draw=color_2!65!black, fill=color_2!30!white] coordinates {  (CO (f64), 23.2883)[44.0]  (CO (f32), 24.3142)[43.0]  (dense (f64), 103.814)[31.1]  (dense (f32), 173.521)[25.5]  (nd6k (f64), 76.6098)[31.5]  (nd6k (f32), 147.247)[35.8]  (average (f64), 22.519412173913043)[35.8]  (average (f32), 31.83350434782609)[34.8]  };
\addplot[draw=color_3!65!black, fill=color_3!30!white] coordinates {  (CO (f64), 23.4426)[44.1]  (CO (f32), 24.2235)[43.9]  (dense (f64), 96.9898)[32.0]  (dense (f32), 168.824)[26.3]  (nd6k (f64), 89.2321)[39.7]  (nd6k (f32), 146.597)[37.8]  (average (f64), 21.41191586956522)[36.9]  (average (f32), 30.375107608695654)[35.2]  };
\addplot[draw=color_4!65!black, fill=color_4!30!white] coordinates {  (CO (f64), 22.765)[44.9]  (CO (f32), 24.9176)[44.9]  (dense (f64), 93.307)[29.9]  (dense (f32), 165.591)[25.9]  (nd6k (f64), 81.355)[40.4]  (nd6k (f32), 140.364)[38.5]  (average (f64), 19.679669347826085)[38.5]  (average (f32), 30.078147173913045)[36.3]  };

 \legend{SPC5 $\beta$(1$\cdot$VS), SPC5 $\beta$(2$\cdot$VS), SPC5 $\beta$(4$\cdot$VS), SPC5 $\beta$(8$\cdot$VS)}
 
        \end{axis}
        \end{tikzpicture}
        \caption{Fujitsu-SVE}
        \label{res:perf:par:sve}  
    \end{subfigure}
    \hfill
    \vspace{-3mm}
    \begin{subfigure}[c]{\textwidth}
        \begin{tikzpicture}
        \begin{axis}[
                height=.18\textheight,
                width=\textwidth,
                ybar=0pt,
                ymin=0,
                clip=false,
                axis lines*=left,
                bar width=3.9pt,
                enlarge x limits=0.05,
                legend style={at={(0.95,1.3)}, anchor=east,legend columns=6, font=\footnotesize},
                ylabel={GFlop/s},
                symbolic x coords={CO (f64),CO (f32),dense (f64),dense (f32),nd6k (f64),nd6k (f32),average (f64),average (f32)},
                xticklabels={CO (f64),CO (f32),dense (f64),dense (f32),nd6k (f64),nd6k (f32),average (f64),average (f32)},
                xtick=data,
                nodes near coords,
                point meta=explicit symbolic,
                every node near coord/.append style={font=\footnotesize},
                yticklabel style = {font=\footnotesize,xshift=0.5ex},
                xticklabel style = {font=\footnotesize,yshift=0.5ex},
                ylabel style = {font=\footnotesize},
                xlabel style = {font=\footnotesize},
                every axis legend/.append style={nodes={right}},
                every node near coord/.append style={font=\tiny, rotate=90, anchor=west},
                x tick label style={rotate=50,anchor=east},
                ymajorgrids = true,
            ]
\addplot[draw=color_1!65!black, fill=color_1!30!white] coordinates {  (CO (f64), 23.1631)[18.4]  (CO (f32), 21.2489)[14.8]  (dense (f64), 44.9592)[12.1]  (dense (f32), 20.6617)[2.2]  (nd6k (f64), 61.1834)[20.7]  (nd6k (f32), 92.6355)[14.7]  (average (f64), 18.420171086956522)[17.5]  (average (f32), 28.35386586956521)[15.9]  };
\addplot[draw=color_2!65!black, fill=color_2!30!white] coordinates {  (CO (f64), 21.2287)[13.6]  (CO (f32), 23.8809)[12.7]  (dense (f64), 61.9837)[15.0]  (dense (f32), 37.347)[3.5]  (nd6k (f64), 68.6185)[20.2]  (nd6k (f32), 110.125)[15.1]  (average (f64), 19.359210000000004)[16.8]  (average (f32), 29.335192391304343)[14.9]  };
\addplot[draw=color_3!65!black, fill=color_3!30!white] coordinates {  (CO (f64), 23.0423)[13.6]  (CO (f32), 24.0104)[13.0]  (dense (f64), 81.9935)[19.6]  (dense (f32), 74.9685)[6.8]  (nd6k (f64), 72.539)[21.1]  (nd6k (f32), 112.158)[14.8]  (average (f64), 20.01204673913043)[17.5]  (average (f32), 30.663484565217384)[15.6]  };
\addplot[draw=color_4!65!black, fill=color_4!30!white] coordinates {  (CO (f64), 21.7373)[13.6]  (CO (f32), 21.8872)[11.8]  (dense (f64), 90.5816)[20.3]  (dense (f32), 112.955)[10.2]  (nd6k (f64), 66.3241)[19.6]  (nd6k (f32), 109.466)[15.3]  (average (f64), 19.241509130434782)[18.1]  (average (f32), 28.986500434782606)[16.0]  };

        \end{axis}
        \end{tikzpicture}
        \caption{Intel-AVX512}
        \label{res:perf:par:avx}        
    \end{subfigure}
        \caption{Performance in Giga Flop per second for parallel computation in double and single precision for our SPC5 kernels on Fujitsu-SVE and Intel-AVX512 architectures.
                 Speedup of parallel SPC5 is computed against the same sequential version and written above the bars.
                 We provide the results for the CO, dense and nd6k matrices, and the average based on all the matrices from the test set.
                 }
        \label{res:perf:par}
\end{figure}
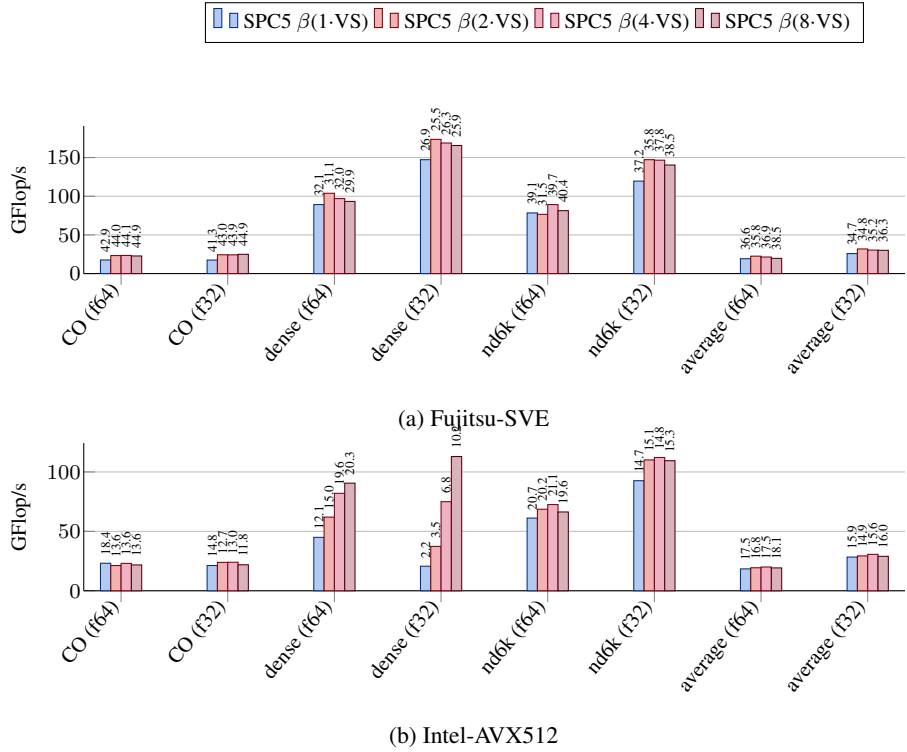